\documentclass{aa}  
\usepackage{graphicx}       
\usepackage{natbib}

\newcommand{\Mo} {$M_{\odot}$}
\newcommand{\Zo} {Z$_{\odot}$}
\newcommand{\Hb} {H$\beta$}
\newcommand{\Ha} {H$\alpha$}
\newcommand{\nodata}{...}
\begin{document}
   \title{Star formation and stellar populations in the Wolf-Rayet(?) luminous compact blue galaxy IRAS %%@
08339+6517\thanks{Based on observations made with NOT (Nordic Optical Telescope) operated on the island of La Palma %%@
by Instituto de Astrof\'\i sica de Canarias in the Spanish Observatories of Roque de Los Muchachos of the Instituto %%@
de Astrof\'\i sica de Canarias.}}

   %\subtitle{I. Overviewing the $\kappa$-mechanism}

   \author{\'Angel R. L\'opez-S\'anchez
          \inst{1},
		  C\'esar Esteban\inst{1}
          \and
          Jorge Garc\'{\i}a-Rojas\inst{1}%\fnmsep\thanks{Just to show the usage
          %of the elements in the author field}
          }

   \offprints{\'Angel R. L\'opez-S\'anchez, \email{angelrls@iac.es}}

\institute{Instituto de Astrof{\'\i}sica de Canarias, E-38200, La Laguna, Tenerife, Spain}

       %       \email{wuchterl@amok.ast.univie.ac.at}
       %  \and
       %      University of Alexandria, Department of Geography, ...\\
       %      \email{c.ptolemy@hipparch.uheaven.space}
       %      \thanks{The university of heaven temporarily does not
       %              accept e-mails}
       %      }

   \date{Received March 23, 2005; Accepted November 23, 2005}

% \abstract{}{}{}{}{} 
% 5 {} token are mandatory
 
  \abstract{IRAS 08339+6517 is a luminous infrared and Ly$\alpha$-emitting starburst galaxy that possesses a dwarf %%@
companion object at a projected distance of 56 kpc. An \ion{H}{i} tidal tail has recently been detected between both %%@
galaxies, suggesting that about 70\% of the neutral gas has been ejected from them. We present deep broad-band %%@
optical images, together with narrow band H$\alpha$ CCD images, and optical intermediate-resolution spectroscopy of %%@
both galaxies. The images reveal interaction features between both systems and strong H$\alpha$ emission in the inner %%@
part of IRAS 08339+6517. The chemical composition of the ionized gas of the galaxies is rather similar. The analysis %%@
of their kinematics also indicates interaction features and reveals an object that could be a candidate tidal dwarf %%@
galaxy or a remnant of an earlier merger. Our data suggest that the \ion{H}{i} tail has been mainly formed from %%@
material stripped from the main galaxy. We find weak spectral features that could be attributed to the presence of %%@
Wolf--Rayet stars in this starburst galaxy and estimate an age of the most recent burst of around 4 -- 6 Myr. A more %%@
evolved underlying stellar population, with a minimal age between 100 -- 200 Myr, is also detected and fits an %%@
exponential intensity profile. A model which combines 85\% young and 15\% old populations can explain both the %%@
spectral energy distribution and the \ion{H}{i} Balmer and \ion{He}{i} absorption lines presented in our spectrum. %%@
The star formation rate of the galaxy is consistently derived using several calibrations, giving a value of $\sim$9.5 %%@
\Mo\ yr$^{-1}$. IRAS 08339+6517 does satisfy the criteria of a luminous compact blue galaxy, rare objects in the %%@
local universe but common at high redshifts, being a very interesting target for detailed studies of galaxy evolution %%@
and formation.}

\titlerunning{SF and stellar populations in the LCBG IRAS 08339+6517}
%%%%%\titlerunning{Interaction and WR features in the LCBG IRAS 08339+6517}
\authorrunning{L\'opez-S\'anchez, Esteban \& Garc\'{\i}a-Rojas}

%\keywords{galaxies: starburst --- galaxies: interactions --- galaxies: abundances --- galaxies: 
%kinematics and dynamics --- stars: Wolf-Rayet --- galaxies: individual: IRAS 08339+6517, 2MASX J08380769+6508579}

   \maketitle
%
%________________________________________________________________

\section{Introduction}

Determining the star formation history in galaxies is fundamental for the understanding of their evolution. In nearby %%@
galaxies, such as those belonging to the Local Group, it is performed studying the stellar content via %%@
color-magnitude diagrams (CMD) and stars as age tracers [see \citet{Grebel99} and references there in]. However, in %%@
distant galaxies it is only interpreted by means of spectral synthesis techniques. The analysis of the interstellar %%@
medium (ISM) in and around galaxies complements the study of their star formation histories. 
Star formation activity is stronger in starburst galaxies, where the dominant and very young stellar population could %%@
even hide the more evolved stellar population presented in it. The more extreme cases are the blue compact dwarf %%@
(BCDs) galaxies, that have very low metal abundances and exhibit a global starburst activity. Although the majority %%@
of these systems possesses an old underlying population with ages of several Gyrs \citep{LT86}, a few of them do not %%@
present it, suggesting that they are really young galaxies \citep{IT99}. The best example of this kind of BCDs is I %%@
Zw 18, a bona fide young galaxy in the local universe \citep{IT04}. At intermediate and high redshifts a %%@
heterogeneous class of vigorous starburst systems with luminosities around $L^*$ ($L^{\star} = 1.0 \times 10^{10} %%@
L_{\odot}$) are observed \citep{GOK03}. They are designated as luminous compact blue galaxies (LCBGs) and their %%@
evolution and nature are still open questions, being fundamental to get a sample of them in the local universe to %%@
investigate the origin of their activity \citep{WJS04}.   

An important subset of starbursts are the so-called Wolf--Rayet (WR) galaxies, whose integrated spectra have a broad %%@
emission feature around 4650 \AA\ that has been attributed to WR stars. This feature consists of a blend of %%@
\ion{N}{iii} $\lambda$4640, \ion{C}{iii}/\ion{C}{iv} $\lambda$4650 and \ion{He}{ii} $\lambda$4686 emission lines %%@
\citep{C91}, the last one being the most prominent line. The WR feature indicates the presence of a substantial %%@
population of these kinds of massive stars whose ages are less than 6 Myr and offers the opportunity to study very %%@
young starbursts \citep{SV98}. Furthermore, they constitute the best direct measure of the upper end of the initial %%@
mass function (IMF), a fundamental ingredient for studying unresolved stellar populations \citep{SGIT00,PSGD02}. %%@
Studying a sample of WR galaxies, \citet{ME00} suggested that interactions with or between dwarf objects could be the %%@
main star formation triggering mechanism in dwarf galaxies and noted that the interacting and/or merging nature of WR %%@
galaxies can be detected only when deep and high-resolution images and spectra are available. The compilation of WR %%@
galaxies performed by \citet{SCP99} lists 139 members, but since then this number has increased %%@
\citep{PH00,BO02,CTS02,PSGD02,LTD03,Tran03,FCCG04,IPG04,PKP04,Kniazev04}, these galaxies have even been detected at %%@
high $z$ \citep{VMCGD04}. In this paper, we add a new member to the list of probable WR galaxies: the luminous %%@
infrared galaxy IRAS 08339+6517, which, furthermore, could be also classified as a luminous compact blue galaxy %%@
(LCBG).

%This assumption seems to be confirmed with new data obtained for our group
%We are developing such analysis and finding increasing evidences of the connection between the detection of WR %%@
%features and the existence of interaction signatures in these kinds of starbursting galaxies %%@
%\citep{LSE03,LSER04a,LSER04b}. 
%detailed analysis of Wolf-Rayet galaxies using both deep images and spectra in optical wavelenghts.

%For example, Schaerer, Guseva, Izotov, Thuan 2000 use the 
%WR feature to constrain $M_{up}$, finding that the observational 
%data are compatible with a Salpeter IMF extending to masses 
%$M_{up} \geq$ 40 \Mo, although Pindao, Schaerer, Rosa 2002 u
%sing VLT data derive a lower limit of 60-90 \Mo\ for $M_{up}$ )

IRAS 08339+6517 is at 80 Mpc (at that distance, 1$\arcsec$ = 388 pc, assuming H$_0$=75 km s$^{-1}$
Mpc$^{-1}$) and was firstly reported in the IRAS Point Source Catalog (1986). \citet{MAM88} performed a %%@
multi-wavelength study and described the object as an exceptionally bright and compact starburst nucleus. %%@
\citet{GDL98} presented a detailed UV analysis, suggesting that O stars must be present in the ionizing cluster(s) of %%@
the galaxy in order to explain the blue wing of the \ion{C}{iv} $\lambda$1550 and \ion{Si}{iv} $\lambda$1400 stellar %%@
absorption lines, although these stars cannot be more massive than about 40 \Mo. \citet{KMH98} noted a P Cygni %%@
profile in the Ly$\alpha$ emission, which also showed two components indicating the chaotic structure of the %%@
interstellar medium in this object. \citet{MHK03} included IRAS 08339+6517 in their $HST$ UV and optical study of %%@
Ly$\alpha$ starbursts. 

In a very recent paper, \citet{CSK04} presented  VLA \ion{H}{i} imaging of this starburst galaxy and found an %%@
extended tidal structure in neutral hydrogen indicating that it is interacting with a nearby companion, 2MASX %%@
J08380769+6508579 (which we designate as the companion galaxy throughout this paper). This feature provides an %%@
evidence for tidally induced starburst episodes. \citet{CSK04} estimated neutral hydrogen masses of $(1.1\pm %%@
0.2)\times 10^9$ \Mo\ and $(7.0\pm 0.9)\times 10^8$ \Mo\ for IRAS 08339+6517 and its companion galaxy, respectively, %%@
and a mass of $(3.8\pm 0.5)\times 10^9$ \Mo\ in the tidal material between them. Consequently, $\sim$ 70\% of the %%@
neutral gas has been removed from one or both galaxies.

We are developing a detailed analysis of Wolf-Rayet galaxies using both deep images and spectra in optical %%@
wavelenghts. Our data show increasing evidences of the connection between the detection of WR features and the %%@
finding of interaction signatures in this kind of starbursts \citep{ME00,LSE03,LSER04a,LSER04b,LS06}. The %%@
characteristics previously observed in IRAS 08339+6517 (strong \Ha\ emission, starburst nature and the detection of %%@
an extended \ion{H}{i} tidal tail in the direction of its companion dwarf galaxy) make it an ideal target to look for %%@
the existence of WR stars. In this way, we have performed deep broad-band optical and narrow-band H$\alpha$ CCD %%@
images together optical intermediate-resolution spectroscopy of IRAS 08339+6517 and its dwarf companion galaxy. Our %%@
main aim is to study their morphology and stellar populations, as well as the distribution, che\-mi\-cal composition %%@
and kinematics of the ionized gas, and to check if WR stars are presented in their youngest bursts.

\begin{figure*}[ht]
\includegraphics[angle=90,width=1\linewidth]{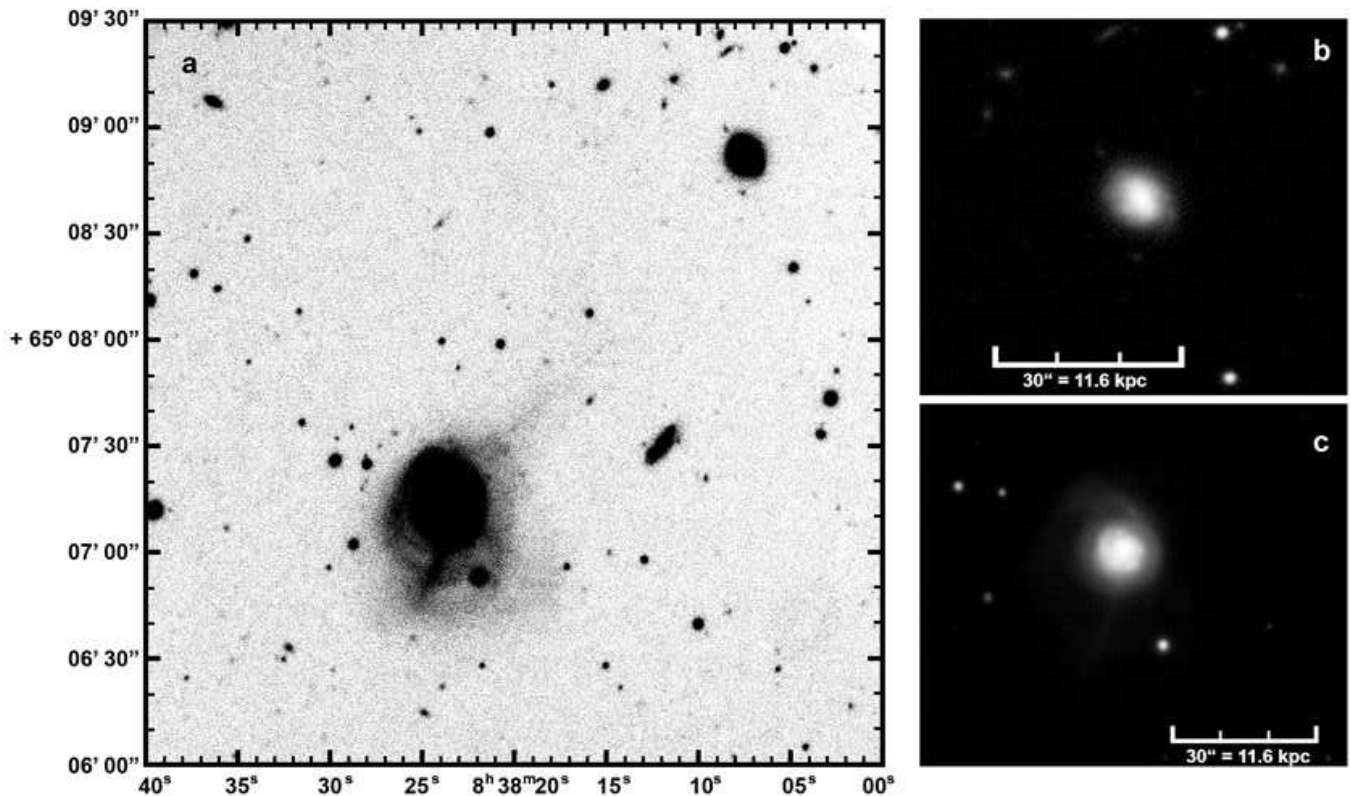}
\caption{\small{(a) Deep optical image of IRAS 08339+6517 and its companion galaxy, 2MASX J08380769+6508579, in $R$. %%@
It has been saturated to reveal the weakest features. (b) The companion galaxy and (c) IRAS 08339+6517 in non-satured %%@
$R$ images. The grayscale is in logarithmic scale in all three cases.}}
\label{figR}
\end{figure*}

\section{Observations and data reduction}

The main data were obtained on 2004 March 20 at the 2.56 m Nordic Optical Telescope (NOT) at Roque de los Muchachos %%@
Observatory (La Palma, Canary Islands, Spain). We used the ALFOSC (Andaluc\'{\i}a Faint Object Spectrograph and %%@
Camera) instrument in both image and spectrograph mode, with a Loral/Lesser CCD detector (2048 $\times$ 2048 pixels) %%@
with a pixel size of 13.5 $\mu$m and spatial resolution of 0.19$\arcsec$ pixel$^{-1}$. Additional $U$ and $B$ images %%@
were obtained on 2005 April 03 using the same telescope and instrument. The journal of all the imaging and %%@
spectroscopical observations can be found in Table~\ref{table1}.

\begin{table}[t!]\centering
  \caption{Summary of observations performed using ALFOSC in 2.56m NOT, with a spatial resolution of 0.19$\arcsec$ %%@
pix$^{-1}$. The position angle, spectral resolution and wavelength range of the intermediate resolution spectroscopy %%@
is shown in the last line.}
  \smallskip
  \label{table1}  
  \small
  \begin{tabular}{ccccc}
    \hline\hline
	\noalign{\smallskip}
    Observations  & Date & Exp. Time (s)   & Filter \\ 
        
	\hline
	\noalign{\smallskip}
	Broad-band &  05/04/03 & 3 $\times$ 300 &  $U$   \\
	imaging    &  05/04/03 & 3 $\times$ 300 &  $B$  \\
               &  04/03/20 & 2 $\times$ 300 &  $V$  \\
	           &  04/03/20 & 3 $\times$ 300 &  $R$  \\
	\noalign{\smallskip}
	Narrow-band&    04/03/20 & 3 $\times$ 300 &  $H\alpha$ (F.19)  \\
	imaging    &    04/03/20 & 2 $\times$ 300 &  $H\alpha$ cont. (F.20)\\
	\noalign{\smallskip}

     Intermediate & 04/03/20 & 3 $\times$ 900 & grism \#7 \\
	 resolution   &  P.A. & $R$ & $\Delta\lambda$  \\
	 spectroscopy & 138${^\circ}$ & 1.5 \AA\ pix$\rm^{-1}$  & 3600-6800 \AA \\ 
    \noalign{\smallskip}
	\hline\hline
  \end{tabular}
\end{table}

\subsection{Optical imaging}

We obtained three 300 second exposures for $U$, $B$ and $R$ standard Johnson filters and two 300 second exposures for %%@
the $V$ filter that were combined to obtain the final images with a PSF FWHM of 0.8$\arcsec$. The standard starfield %%@
SA107-602 \citep{L92} was used to flux calibrate the final images. It was observed at different airmasses during the %%@
night. Twilight images of different zones of the sky (blank fields) were taken for each filter in order to perform %%@
the flat-field correction. All the reduction process (bias subtraction, flat-fielding and flux calibration) was done %%@
with IRAF\footnote{IRAF is distributed by NOAO which is operated by AURA Inc., under cooperative agreement with NSF} %%@
package following standard procedures. IRAF software was also used to determine the 3$\sigma$ contours over the sky %%@
background level and to obtain the photometric values of the galaxies.

\subsection{H$\alpha$ imaging}

The narrow-band filters for H$\alpha$ and red-continuum ima\-ges (centered at 6687 and 6571 \AA\ and with a FWHM of %%@
50 and 47 \AA, respectively) were selected taking into account the recession velocity of the object given by the  %%@
NASA/IPAC Extragalactic Database (NED). Three (two for the conti\-nuum) exposures of 300 seconds were added to obtain %%@
the final image in each filter. The reduction process was similar to the broad-band images. The absolute flux %%@
calibration was achieved by taking short exposures of the spectrophotometric standard star Feige~56 \citep{Hamuy92} %%@
at different airmasses. 

To obtain the continuum-subtracted \Ha\ image, we followed the standard procedure given by \citet{Bernabe03}. A %%@
scaling factor between the \Ha\ and the continuum frames was determined using non-saturated field stars (in our case, %%@
this scaling factor was 0.9). The FWHM of the PSF for both images was similar (around 0.6$\arcsec$), so no Gaussian %%@
filter was applied. The finally flux-calibrated and continuum-subtracted \Ha\ image was then produced by subtracting %%@
the scaled continuum frame from the \Ha\ frame. The W(\Ha) image was also obtained following the standard procedure, %%@
correcting the W(\Ha) for the underlying stellar population of the disk.

\subsection{Intermediate-resolution spectroscopy}

We used ALFOSC in spectrographic mode with a slit of 6.4$\arcmin$ (the field of view of the camera) long and %%@
1$\arcsec$ wide to obtain intermediate-resolution spectroscopy of the galaxies. Grism \#7 with 600 rules mm$^{-1}$ %%@
and resolution of 1300 was used, with a dispersion of 111 \AA\ mm$^{-1}$ and spectral resolution of 1.5 \AA\ %%@
pix$^{-1}$. Each spectrum spans the wavelength range 3200 to 6800 \AA\ (redshift corrected). The slit position was %%@
located at P.A. = 138${^\circ}$, which was chosen in order to cross the center of IRAS 08339+6517 and its companion %%@
galaxy. Three 15 minute exposures were taken and combined to obtain good signal-to-noise and an appropriate removal %%@
of cosmic rays in the final spectra. We used comparison He and Ne lamp exposures for the wavelength calibration of %%@
the spectra. The absolute flux calibration was achieved by observations of the standard star Feige 56 \citep{M88}. %%@
The observations were made at air masses between 1.1 and 1.3. Consequently, no correction was made for atmospheric %%@
differential refraction.   

\begin{table}[t]\centering
 \caption{General properties of the galaxies}
 \label{table2}
 \scriptsize
 \begin{tabular}{lcc}
   \\
   \hline\hline
   \noalign{\smallskip}
    Property  & IRAS 08339+6517 & companion  \\
   \hline
    \noalign{\smallskip}
      M$_B$$^0$ &  $-$21.57 $\pm$ 0.04 &  $-$18.21 $\pm$ 0.04  \\
      m$_B$$^0$ &     12.94 $\pm$ 0.04 &     16.31 $\pm$ 0.04  \\
L$_B/L_{\odot}$ &  (6.69 $\pm$ 0.24) $\times\ 10^{10}$ & (2.99 $\pm$ 0.11) $\times\ 10^{9}$  \\
 E$(B-V)$$\rm^a$&      0.16 $\pm$ 0.02 &      0.13 $\pm$ 0.02  \\
  \noalign{\smallskip}
        $(U-B)$$^0$&   $-$0.51 $\pm$ 0.08 &   $-$0.16 $\pm$ 0.10  \\
	    $(B-V)$$^0$&      0.02 $\pm$ 0.08 &      0.20 $\pm$ 0.08  \\
        $(V-R)$$^0$&      0.24 $\pm$ 0.08 &      0.26 $\pm$ 0.08  \\
  $(V-J)^{0,\rm b}$&      1.39 $\pm$ 0.06 &      1.56 $\pm$ 0.12 \\ 
  $(J-H)^{0,\rm b}$&      0.64 $\pm$ 0.05 &      0.21 $\pm$ 0.25  \\
$(H-K_s)^{0,\rm b}$&      0.23 $\pm$ 0.06 &      0.68 $\pm$ 0.28  \\
   
 \noalign{\smallskip}
 \noalign{\smallskip}
 
 M$_{\ion{H}{ii}}$ (M$_{\odot}$)        & (1.45 $\pm$ 0.07)$\times 10^7$ & (2.6 $\pm$ 0.3)$\times 10^5$ \\
 M$_{\ion{H}{i}}$ (M$_{\odot}$) $\rm^c$ & (1.1 $\pm$ 0.2)$\times 10^9$ & (7.0 $\pm$ 0.9)$\times 10^8$ \\
 M$_{dust}$   (M$_{\odot}$)$\rm^d$      &  4.5$\times 10^6$      & \nodata  \\  
 M$_{Kep}$ (M$_{\odot}$)                & (10 $\pm$ 3)$\times 10^9$    & (8 $\pm$ 2)$\times 10^9$ \\
 M$_{Kep}$/M$_{\ion{H}{i}}$             &  9.1                   & 1.1 \\
 M$_{\ion{H}{ii}}$/M$_{\ion{H}{i}}$     &  0.013                 & 0.0004 \\
 L$_B$/M$_{Kep}$                        &  6.7                   & 0.37 \\   
 L$_B$/M$_{H\,I}$                       &  60.8                  & 4.2 \\
 L$_B$/M$_{dust}$                       &  1.5$\times 10^4$      & \nodata \\ 
 \noalign{\smallskip}
 \noalign{\smallskip}
% T$_{dust}$   (K)                       &  46                    & \nodata \\
 
% \noalign{\smallskip}
%  \noalign{\smallskip}
  SFR (\Mo\ yr$^{-1}$)                 &   9.5                  & 0.17 \\
  $\tau_{gas}$    (Gyr)                &  0.15                  & 5.4 \\
  Age $\rm^e$ (Myr)  &            4 -- 5     &       5 -- 6           \\
  Age $\rm^f$ (Myr) &            30 -- 50     &       $>$ 250         \\
$\Delta$v$_r$$\rm^g$  (km s$^{-1}$)  &    0   &    20 $\pm$ 10  \\
12+log O/H$\rm^h$          &8.45 $\pm$ 0.10 &   8.38 $\pm$ 0.10 \\
\noalign{\smallskip}
  \noalign{\smallskip}
  N(WR)$\rm^i$             & $\sim$ 310   & \nodata \\
  WR/(WR+O)$\rm^i$                & 0.03       & \nodata \\               
 \noalign{\smallskip}
 \hline\hline
 \end{tabular}
 \begin{flushleft}
 $\rm^a$ Derived from C(H$\beta$). See \S 3.1 for more details. \\
 $\rm^b$ Derived from the values of 2MASS shown by the NED.\\
 $\rm^c$ From \citet{CSK04}.\\
 $\rm^d$ Derived using the \citet{HSH95} relation between the 60 and 100 $\mu$m FIR fluxes and the dust mass (last %%@
equation in p. 678).\\
 $\rm^e$ Estimated age of the most recent bursts using our \Ha\ and spectroscopical data. See \S 4.1 for more %%@
details.\\
 $\rm^f$ Estimated age comparing our optical colors with the STARBURSTS 99 \citep{L99} and PEGASE.2 \citep{PEGASE97} %%@
models. See \S 4.1 for more details.\\
 $\rm^g$ Radial velocity with respect to the center of IRAS 08339+6517.   \\
 $\rm^h$ Derived using the \citet{P01} empirical calibration (see \S3.4.2).\\
 $\rm^i$ Approximate number of WR stars and WR/(WR+O) ratio (see \S4.3).\\
 \end{flushleft}
\end{table} 

All the CCD frames for spectroscopy were reduced using standard IRAF procedures to perform bias correction, %%@
flat-fielding, cosmic-ray rejection, wavelength and flux calibration, and sky subtraction. The correction for %%@
atmospheric extinction was performed using an average curve for the continuous atmospheric extinction at Roque de los %%@
Muchachos Observatory. For the two-dimensional spectra two different apertures were defined along the spatial %%@
direction to extract the final one-dimensional spectra of the galaxies, centering each aperture on its brightest %%@
zone. Small two-dimensional distortions were corrected by fitting the maxima of [\ion{O}{ii}] %%@
$\lambda\lambda$3726,3729 doublet and H$\alpha$ emission lines (between 2 and 3 pixels, 0.4 -- 0.6$\arcsec$).

\section{Results}

\subsection{The extinction toward IRAS 08339+6517}

We have carefully studied the reddening coefficient, C(\Hb), derived from our spectra using the Balmer decrement (see %%@
\S3.4) in order to achieve accurate photometric and spectroscopic data of IRAS 08339+6517 and its companion galaxy. %%@
However, the fluxes of nebular Balmer lines in the optical spectra must be also corrected for underlying stellar %%@
absorption. In this way, we performed an iterative procedure to derive the reddening coefficient and the equivalent %%@
widths of the absorption in the hydrogen lines, $W_{abs}$, to correct the observed line intensities for both effects. %%@
We assumed that the equivalent width of the absorption components is the same for all the Balmer lines and used the %%@
relation given by \citet{MB93} to the absorption correction,
\begin{eqnarray}
C(H\beta)=\frac{1}{f(\lambda)} \log\Bigg[\frac{\frac{I(\lambda)}{I(H\beta)}\times %%@
\Big(1+\frac{W_{abs}}{W_{H\beta}}\Big)} {\frac{F(\lambda)}{F(H\beta)}\times %%@
\Big(1+\frac{W_{abs}}{W_{\lambda}}\Big)}\Bigg],
\end{eqnarray}
for each detected hydrogen Balmer line. In this equation, $F(\lambda)$ and $I(\lambda)$ are the observed and the %%@
theoretical fluxes (unaffected by reddening or absorption); $W_{abs},\ W_{\lambda}$ and $W_{H\beta}$ are the %%@
equivalent widths of the underlying stellar absorption, the considered Balmer line and H$\beta$, respectively, and %%@
$f(\lambda)$ is the reddening curve normalized to \Hb\ using the \citet{seaton79} law. We have used the pairs %%@
\Ha/\Hb, H$\gamma$/\Hb\ and H$\delta$/\Hb\ for the main galaxy but only the two first line ratios for the companion %%@
object. We have considered the theoretical ratios between these pairs of \ion{H}{i} Balmer lines expected for case B %%@
recombination given by \citet{SH95} and appropriate electron temperatures around 10$^4$ K and electron densities %%@
around 100 cm$^{-3}$ (\Ha/\Hb\ = 2.86, H$\gamma$/\Hb\ = 0.469 and H$\delta$/\Hb\ = 0.259). The C(\Hb) and $W_{abs}$ %%@
that provide the best match between the corrected and the theoretical line ratios are shown in Table~\ref{table4}.  

Once we have obtained the reddening coefficient, the photometric data can also be corrected for extinction by %%@
interstellar dust. We used the standard \citet{seaton79} law and the standard ratio of $R_V$ = $A_V$/$E(B-V)$ = 3.1 %%@
to obtain the extinction in $V$, $A_V \sim$ 2.11 $\times$ C(H$\beta$) and the color excess, $E(B-V)$ = 0.68 $\times$ %%@
C(H$\beta$). The data in $U$, $B$ and $R$ were corrected using $A_U$ = 1.531 $\times$ $A_V$, $A_B$ = 1.324 $\times$ %%@
$A_V$ and $A_R$ = 0.748 $\times$ $A_V$ \citep{RL85}, whereas the data in \Ha\ filter were corrected using %%@
$A_{H\alpha}$ = 1.61 $\times$ C(H$\beta$) \citep{seaton79}. 

As we explain in \S3.4, we have extracted eight different apertures to study the ionized gas in IRAS 08339+6517. The %%@
C(H$\beta$) derived varies between 0.18 and 0.30, implying a color excess between 0.12 and 0.20. We have adopted the %%@
average C(H$\beta$) (= 0.24) of these eight zones as representative for the galaxy. The derived color excess used to %%@
correct the photometric data is $E(B-V)$ = 0.16 (see Table~\ref{table2}).

\citet{SFD98} presented an all-sky map of infrared dust emission, determining a color excess $E(B-V)$ = 0.092 %%@
(assuming $R_V$ = 3.1) at the Galactic longitude and latitude of IRAS 08339+6517. This value is lower than the %%@
$E(B-V)$ we obtain for the galaxies (0.16 and 0.13), indicating that a considerable fraction of the dust in the line %%@
of sight is external to the Milky Way, especially in the main object.

%Using the dust to gas ratio given by Stark et al. (1992), 
%a color excess of 0.41 is derived, a factor of 2.6 higher 
%than the one we have derived from the Balmer decrement.

%Nosotros usamos la galactica, quizas la relacion entre 
%N y E es DISTINTA en otros objetos, 
%la proporcion seria 1.43 en lugar de 2.9, casi la mitad.
%BUSCAR: dust to gas ratio, segun IV97, the external extinction 
%can be estimated from the neutral hydrogen column density 
%if a metallicity-dependent dust-to-gas ratio is assumed and the 
%extinction is assumed to occur principally outside the H II regions.

%***se puede derivar E(B-V) de la densidad columnar en radio, 
%$E(B-V)= 2.9 \times 10^{-22} N_{H\,I}$  *** Preguntar a Hibbard o Lourdes??.
% Esta relacion es la usada normalmente para la Galaxia 
%(Stark et al. 1992,ApJS 79,77). En el Binney viene esta: 
%$E(B-V)= 1.7 \times 10^{-22} N_{H\,I}$  sale 0.26 usando mapa 
%radio Canon04, n=15E20, relacion para la Galaxia que viene en  Kent, Dame \& 
%Fazio (1991, ApJ 378, 131). COn esta, el modelo de r-X daria E=0.24

\begin{figure}[t!]
\centering
\includegraphics[width=0.9\linewidth]{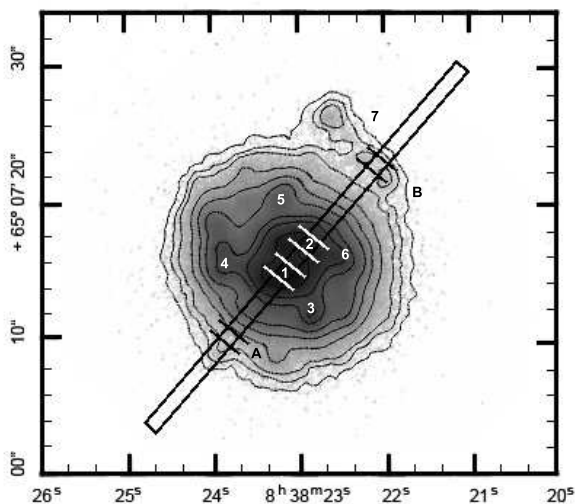}
\caption{\small{Continuum-subtracted \Ha\ image of IRAS 08339+6517. The contours represent 3$\sigma$, 5$\sigma$, %%@
7$\sigma$, 15$\sigma$, 30$\sigma$, 70$\sigma$, 140$\sigma$, 230$\sigma$, 430$\sigma$ and 700$\sigma$ above the sky %%@
level. We also show the position of the slit used for spectroscopy, with P.A. 138${^\circ}$, and label the different %%@
individual regions that we have analyzed.}}
\label{figHa}
\end{figure}

The importance of the extinction correction was previously remarked on by \citet{GDL98}, who studied the variation of %%@
the color excess derived from various methods: the Balmer decrement (that gave an $E(B-V) = 0.52$, a similar value %%@
that the one derived by \citet{MAM88} using the \Ha/\Hb\ ratio, $E(B-V) = 0.55$, although they do not correct for the %%@
underlying stellar absorption) and the change of the UV continuum slope [which indicates lower values for the %%@
extinction, $E(B-V) = 0.1 - 0.2$]. They explain this difference in terms of a nonhomogeneous distribution of the %%@
dust, although it could also be due to the variation of the evolutionary stage of the stellar population with the %%@
aperture size. The data derived from our optical spectra suggest that $E(B-V)$ is not so high as those derived by %%@
\citet{GDL98} and \citet{MAM88} using the same method but is similar to the value derived from the UV data. This %%@
result is supported by \citet{SS98}, who studied $ROSAT\ PSPC$ sere\-di\-pi\-tious observations of IRAS 08339+6517 %%@
and concluded that the spectral fitting yields an X-ray temperature of $kT$ = 0.58 keV, with $Z$ = 0.02$Z_{\odot}$ %%@
and a column of 1.40 $\times$ 10$^{21}$ cm$^{-2}$, implying substantial local absorption. Using the dust-to-gas ratio %%@
given by \citet{KDM91}, a color excess of 0.24 is derived, a bit higher than the one we have assumed from the Balmer %%@
decrement, but not as high as that derived by \citet{GDL98}.

However, we have noted significant differences in C(\Hb) derived for different zones and apertures inside IRAS %%@
08339+6517. We briefly discuss this in \S4.4.

\begin{table*}[t!]\centering
 \caption{\Ha\ properties of the galaxies and studied knots inside IRAS 08339+6517}
 \label{table3}
 \scriptsize
 \begin{tabular}{lcccccc}
   \\
   \hline\hline
   \noalign{\smallskip}
  Knot  & Flux (10$^{-13}$erg cm$^{-2}$ s$^{-1}$)& Luminosity (10$^{41}$erg s$^{-1}$)&  SFR (\Mo\ yr$^{-1}$)  & %%@
M$_{\ion{H}{ii}}$ (10$^6$\Mo) & $-W$(\Ha) (\AA) & Age (Myr) \\
   \hline
    \noalign{\smallskip}
IRAS 08339+6517& 16.0 $\pm$ 0.7& 12.0 $\pm$ 0.6 & 9.5 $\pm$ 0.5 & 14.5 $\pm$ 0.7 &110 $\pm$ 10& 4.4 $\pm$ 0.2\\ 
  \noalign{\smallskip}
  \# 1         &  4.3 $\pm$ 0.3&  3.3 $\pm$ 0.2 &2.62 $\pm$ 0.18&  4.1 $\pm$ 0.3 & 150 $\pm$ 15& 3.3 $\pm$ 0.2 \\
  \# 2         &  2.3 $\pm$ 0.2& 1.74 $\pm$ 0.14&1.38 $\pm$ 0.11& 2.09 $\pm$ 0.17& 103 $\pm$ 9 & 4.4 $\pm$ 0.2 \\ 
  \# 3         &0.47 $\pm$ 0.05& 0.36 $\pm$ 0.04&0.28 $\pm$ 0.03& 0.43 $\pm$ 0.04& 129 $\pm$ 13& 3.9 $\pm$ 0.2 \\ 
  \# 4         &0.76 $\pm$ 0.08& 0.57 $\pm$ 0.06&0.45 $\pm$ 0.05& 0.69 $\pm$ 0.07& 101 $\pm$ 10& 4.4 $\pm$ 0.2 \\ 
  \# 5         &0.24 $\pm$ 0.03& 0.18 $\pm$ 0.02&0.14 $\pm$ 0.02& 0.22 $\pm$ 0.03&  95 $\pm$ 11& 4.5 $\pm$ 0.3 \\ 
  \# 6         &0.70 $\pm$ 0.07& 0.53 $\pm$ 0.05&0.42 $\pm$ 0.04& 0.63 $\pm$ 0.06& 102 $\pm$ 13& 4.4 $\pm$ 0.3 \\ 
  \# 7 (NW arm)&0.096$\pm$ 0.015&0.073$\pm$ 0.011&0.06 $\pm$ 0.01&0.087 $\pm$ 0.013&40 $\pm$ 6 & 5.4 $\pm$ 0.3 \\
  \noalign{\smallskip}
  Companion    &0.28 $\pm$ 0.03& 0.21 $\pm$ 0.02& 0.17 $\pm$ 0.02&0.26 $\pm$ 0.03 & 31 $\pm$ 5 & 5.5 $\pm$ 0.3 \\ 
 \noalign{\smallskip} 
 \hline\hline
 \end{tabular}
\end{table*} 

\begin{figure}[t!]
\includegraphics[angle=270,width=\linewidth]{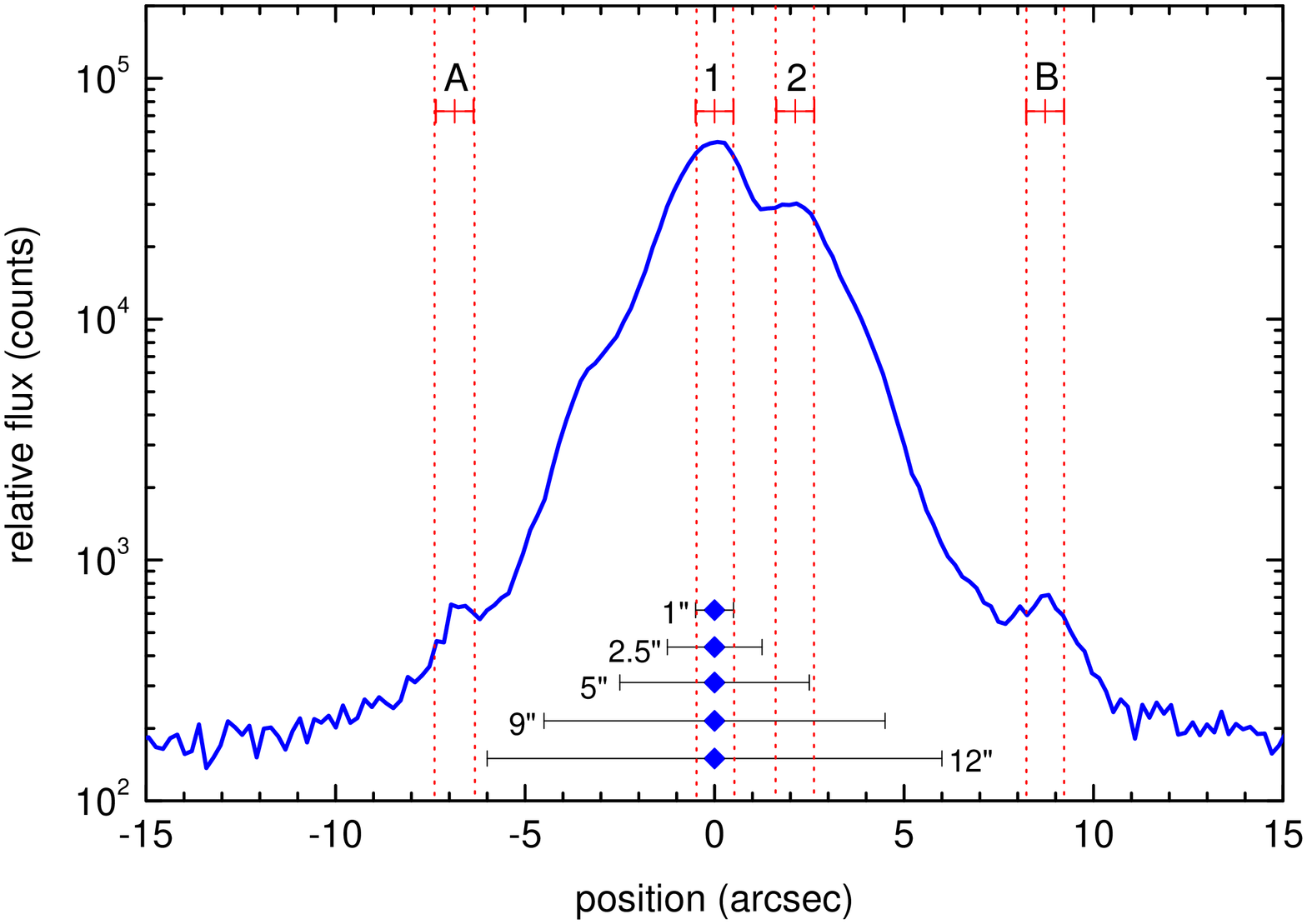}
\caption{\small{Relative \Ha\ flux along the slit (see Figure~\ref{figHa}). The different zones for which we have %%@
extracted spectra are indicated and labeled. Knots \#1 (the center of the galaxy), \#2, A and B are 1$\arcsec$ in %%@
size. We also show four additional apertures extracted to study spatial variations of the ionized gas inside the %%@
galaxy. They have 2.5$\arcsec$, 5$\arcsec$, 9$\arcsec$ and 12$\arcsec$ in size. The aperture of 12$\arcsec$ %%@
represents the integrated spectrum of the galaxy, that we show in Figure~\ref{spectra}(top).}}
\label{rendija}
\end{figure}

\subsection{Optical imaging}

In Figure~\ref{figR}a we show our final deep $R$ image, which has been saturated to show the faintest structures. %%@
This deep ima\-ge reveals a disturbed morphology in the outer regions of IRAS 08339+6517. The most relevant are a %%@
bright ray at the south of the galaxy and a long arc of material connecting the north of the galaxy with the southern %%@
bright ray. This arc seems to be broken into two parts to the SE of the galaxy. Another weaker arc goes from the west %%@
towards the end of the bright ray. Furthermore, a very diffuse plume is found in the NW of the galaxy, precisely in %%@
the direction towards the companion galaxy. This plume coincides with the beginning of the tidal \ion{H}{i} material %%@
discovered by \citet{CSK04} between the two galaxies.

The non-saturated images of the galaxies are shown in Figures~\ref{figR}b and c. IRAS 08339+6517 shows a circular %%@
morphology. Assuming that it has a disk geometry, this fact would indicate that the galaxy is nearly face on. Several %%@
structures are found in its central areas following an apparently symetrical pattern from the nucleus. Although they %%@
are not clearly resolved, they show lower $U-B$ values and possess important \Ha\ emission features (see next %%@
section), indicating that the inner area of the galaxy has relatively high star-formation activity.  These facts %%@
could suggest the presence of a disk-like structure inside IRAS 08339+6517.

The companion galaxy has a not very excentric elliptical morphology, with no relevant feature inside it. It shows %%@
bluer colors in its external areas than in its center.    

We have performed integrated aperture photometry for the galaxies in the optical filters. We have determined the area %%@
for which we have integrated the flux using the 3$\sigma$ level isophote over the average sky level in the $B$ image. %%@
We have used the same area in $U$, $V$ and $R$, correcting all values for reddening as indicated in the previous %%@
section. We have determined the errors for the photometry by con\-si\-de\-ring the FWHM of the PSF, sky level and %%@
flux calibration for each frame.

The final integrated photometric values derived for each galaxy are listed in Table~\ref{table2}. We also show the %%@
NIR colors derived from the values given by 2MASS, that are also corrected by reddening. IRAS 08339+6517 has bluer %%@
colors than its companion object. A detailed analysis of the surface brightness reveals the presence of two different %%@
stellar populations inside IRAS 08339+6517, as we will see in \S4.1. The $B$ magnitudes derived implying an absolute %%@
blue magnitude of $M_B \sim -21.6$ for the main galaxy and $M_B \sim -18.3$ for its companion.

\begin{figure*}[t]
\centering
\includegraphics[angle=270,width=1\linewidth]{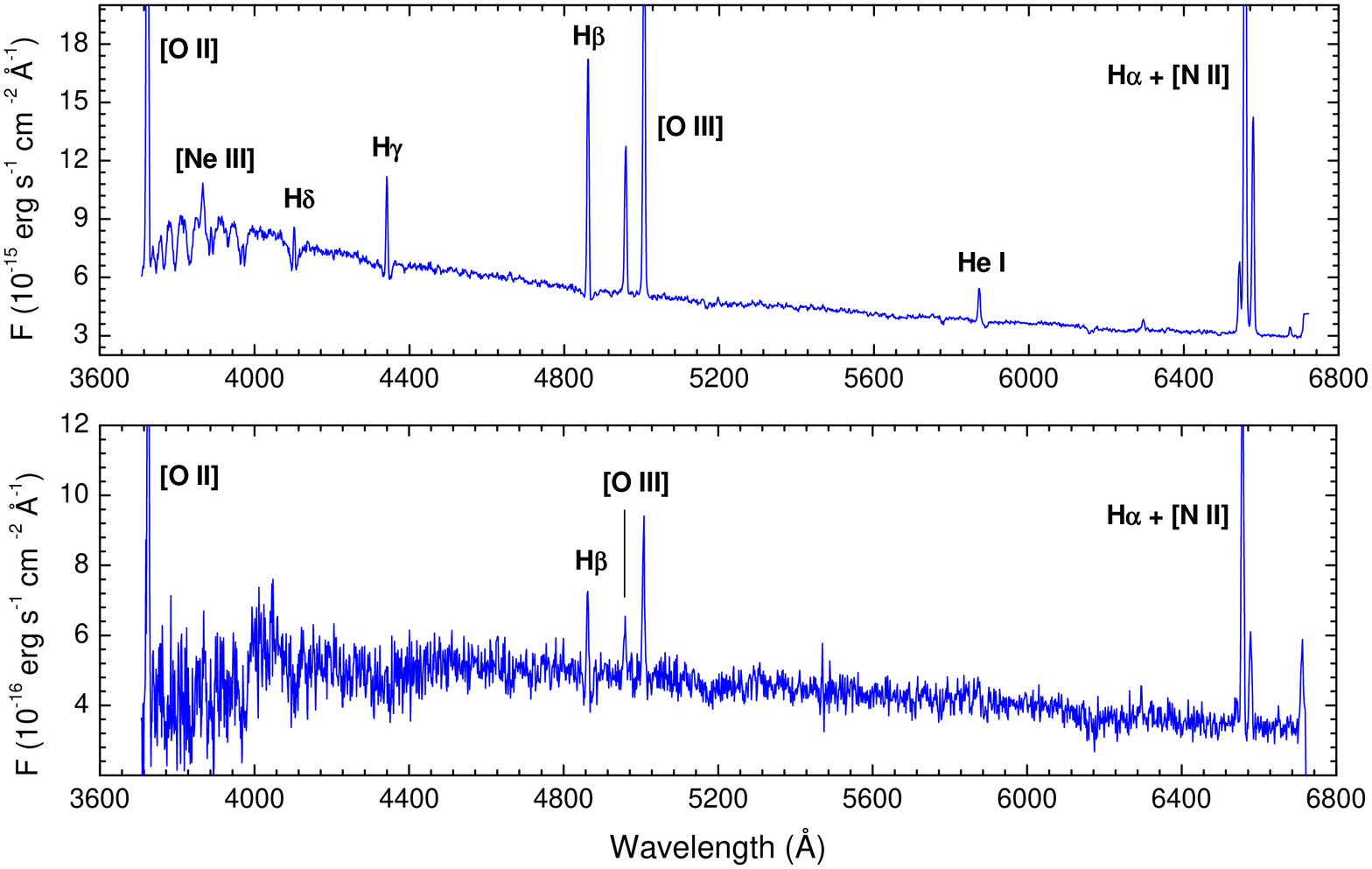}
\protect\caption[ ]{\small{Redshift-corrected spectra of IRAS 08339+6517 (top) and the companion galaxy 2MASX %%@
J08380769+6508579 (bottom). The most important emission lines are labeled. The spectra have been scaled down in flux %%@
in order to distinguish the faint lines. They are not corrected by reddening. Note the high continuum level and the %%@
\ion{H}{i} absorptions due to the underlying stellar population in the spectrum of the main galaxy.}}
\label{spectra}
\end{figure*}

\subsection{\Ha\ imaging}

The \Ha\ map of IRAS 08339+6517 is shown in Figure~\ref{figHa}. The slit position used for spectroscopy, as well as %%@
different regions inside the galaxy that we have also extracted and analyzed, are indicated and labeled. The galaxy %%@
has an almost circular morphology in \Ha\ emission, except for the arm located to its northwest (region \#7 in %%@
Figure~\ref{figHa}). This feature is also in the direction of the companion galaxy. Several star-formation knots are %%@
found inside the disk, the central (region \#1) being  the brightest one. This indicates that the entire galaxy is %%@
undergoing a considerable star-formation activity. Regions \#4 (which seems to be connected to \#1 and is broken into %%@
three small knots) and \#6 (connected to \#2) are elliptic and extend from the center to the outer regions, %%@
suggesting diffuse spiral arms. The NW arm also shows three knots of weak \Ha\ emission. Only the brightest is found %%@
in the optical images, showing $U-B\ \sim -0.4$, $B-V\ \sim 0.1$ and $V-R\ \sim 0.2$ colors. The companion galaxy is %%@
weaker in \Ha\ emission but it also shows star-formation activity. Its morphology is irregular and split into several %%@
weak knots located not in its center but in its outer areas. 

We obtained the \Ha\ fluxes for both galaxies and some regions inside IRAS 08339+6517 using this flux-calibrated %%@
continuum-subtracted \Ha\ image. The boundaries of the zones were determined  selecting each distinct emission peak %%@
as a separate region by eye. We have corrected the \Ha\ fluxes for both reddening (see \S 3.1) and [\ion{N}{ii}] %%@
contamination. The [\ion{N}{ii}] contamination was corrected using the [\ion{N}{ii}] $\lambda$6583/\Ha\ and %%@
[\ion{N}{ii}] $\lambda$6548/\Ha\ ratios derived from our spectra and the transmittance value of the \Ha\ filter for %%@
these wavelengths. In the zones for which we have no spectra we used the average [\ion{N}{ii}]/\Ha\ ratios obtained. %%@
The errors for the \Ha\ flux was determined considering the FWHM of the PSF, sky level, the flux calibration, and the %%@
contributions of the reddening and [\ion{N}{ii}] contamination, which vary between 4\% (for the main galaxy) and 16\% %%@
(for the NW arm). The final fluxes are shown in Table~\ref{table3}. The \Ha\ flux derived for IRAS 08449+6517 is in %%@
excellent agreement with that obtained by \citet{MAM88}, $F_{H\alpha} = 1.3 \times 10^{-12}$ erg s$^{-1}$ cm$^{-2}$.

The luminosity of the galaxies and regions was calculated assuming a distance of 80 Mpc for the system. For %%@
comparison, the giant extragalactic \ion{H}{ii} region 30 Doradus in the LMC has an \Ha\ luminosity of $1.5 \times %%@
10^{40}$ erg s$^{-1}$ \citep{Ken84}, similar to region \#5. Assuming that the ionization is due to hot stars and that %%@
a single O7V star contributes with a luminosity of $L_{H\alpha} = 1.36 \times 10^{36}$ erg s$^{-1}$ \citep{SV98} to %%@
the total flux, we estimated a population of around $\sim$88000 O7V stars for IRAS 08339+6517.

We have derived the star formation rate (SFR) and the mass of the ionized gas, $M_{H II}$, using the standard %%@
\citet{K98} calibration. For IRAS 08339+6517 we find a SFR$_{H\alpha}$ = 9.5 \Mo\ yr$^{-1}$, which is in excellent %%@
agreement with previous studies. We will compare the SFR derived from different wavelengths in \S 4.2. For the mass %%@
of the ionized gas, we have derived a value of $M_{H II} \sim 1.5 \times 10^7$ \Mo.  

We have obtained the equivalent width of the \Ha\ emission, $W($\Ha), of the galaxies and the knots inside IRAS %%@
08339+6517 from the \Ha\ image. The results are compiled in Table~\ref{table3}. All the objects present very similar %%@
$W($\Ha), except the center of the galaxy (knot \#1, which has the highest value) and the companion galaxy and the NW %%@
arm (both showing lowest values). The equi\-va\-lent width of the \Ha\ emission can be used to estimate the age of %%@
the bursts. We have performed this estimation using the STARBURST99 \citep{L99} models for an instantaneous starburst %%@
(see \S4.1 for details). The results for each object are found in Table~\ref{table3}.

Studying the complex structure of the Ly$\alpha$ emission observed in IRAS 08339+6517, \citet{MHK03} conclude that an %%@
expanding superbubble of ionized gas powered by the starburst is moving toward the observer at 300 km s$^{-1}$ with %%@
respect to the central \ion{H}{ii} region. This is supported by the detection of X-ray emission with \emph{Einstein} %%@
\citep{gioia90} and \emph{ROSAT} \citep{SS98} observatories. The strong \Ha\ emission detected throughout the disk of %%@
the galaxy is compatible with this scenario.

\begin{table*}\centering
  \caption{Dereddened line intensity ratios with respect to I(H$\beta$)=100 for the two galaxies and the selected %%@
knots of IRAS 08339+6517. Correction for underlying stellar absorption in \ion{H}{i} Balmer lines has been also %%@
applied.}
  \label{table4}  
  \footnotesize
  \begin{tabular}{lrcccccc}
  \noalign{\smallskip}
    \hline\hline
	\noalign{\smallskip}
	Line     & f($\lambda$)& IRAS 08339+6517      & \#1         &   \#2  &    A  &      B &   Companion  \\
	\hline    
	\noalign{\smallskip}
 3727 $[$O II$]\rm^a$& 0.27  & 288 $\pm$ 11  & 222 $\pm$ 20  &  242 $\pm$ 17  &  322 $\pm$ 115 & 197 $\pm$ 100 & 310 %%@
$\pm$ 44 \\
 3869 $[$Ne III$]$& 0.23  &16.8 $\pm$ 3.2 &16.7 $\pm$ 4.9 & 11.5 $\pm$ 2.4 &	\nodata & \nodata	& 17.6: \\
 3889 He I + H8	  & 0.22  &10.6 $\pm$ 3.1 &5.3:           &   8.5:         &	\nodata & \nodata	& \nodata \\
 3968 $[$Ne III$]$ +H7&0.21&9.7 $\pm$ 0.9 &5.8: 	      &	  4.3: 	       &	\nodata & \nodata	& 6.9: \\
 4101 H$\delta$   & 0.18  &25.9 $\pm$ 3.1 & 25.9 $\pm$ 3.6 & 25.9 $\pm$ 4.6 &	\nodata & \nodata	& 23: \\
 4144 He I        & 0.17  &2.4: &\nodata      &\nodata         &  \nodata & \nodata   & \nodata \\
 4340 H$\gamma$   & 0.135 &46.9 $\pm$ 3.2 &46.9 $\pm$ 3.7 & 46.9 $\pm$ 3.9 &	\nodata & \nodata	& 42: \\
 4363 $[$O III$]$ & 0.13  &0.70:          & 0.8:        &  1.0:          &	\nodata & \nodata	& \nodata \\
% 4414 $[$Fe III$]$& 0.11  &0.65: &\nodata      &\nodata         &  \nodata & \nodata   & \nodata \\
 4471 He I	      & 0.10  &3.17 $\pm$ 0.50& 1.4:   	    &  1.3:		     &  \nodata & \nodata	& \nodata \\
 4658 $[$Fe III$]$& 0.05  &1.20 $\pm$ 0.55& 0.8:	        & 	\nodata      &	\nodata & \nodata	& \nodata \\
 4686 He II       & 0.05  &\nodata        & 1.9:          &\nodata         &  \nodata & \nodata   & \nodata \\
 4861 H$\beta$    & 0.00  & 100 $\pm$ 3 & 100 $\pm$ 4 & 100 $\pm$ 4 & 100 $\pm$ 30  & 100 $\pm$ 25 & 100 $\pm$ 13 \\
 4959 $[$O III$]$ &$-$0.02& 56 $\pm$ 2&  63 $\pm$ 4   &   49 $\pm$ 4	&   34: &  64 $\pm$ 24   &  44 $\pm$ 9 \\
 5007 $[$O III$]$ &$-$0.03& 168 $\pm$ 4& 190 $\pm$ 10  &  163 $\pm$ 9   &  102 $\pm$ 36  & 211 $\pm$ 62   & 135 $\pm$ %%@
19 \\
 5200 $[$N I$]$   &$-$0.05& 2.38 $\pm$ 0.16& 2.1:           &	2.3: 	     &  \nodata & \nodata	& \nodata \\
 
 5516 $[$Cl III$]$&$-$0.15&0.6: &\nodata      &\nodata         &  \nodata & \nodata   & \nodata \\
 5538 $[$Cl III$]$&$-$0.16&0.5: &0.7:      &\nodata         &  \nodata & \nodata   & \nodata \\
 
 5755 $[$N II$]$  &$-$0.21& 0.82 $\pm$ 0.29&1.1:   &   0.6:	     &  \nodata & \nodata	& \nodata \\
 5876 He I	      &$-$0.23& 11.6 $\pm$ 0.8 &13.3 $\pm$ 1.8 & 11.7 $\pm$ 1.9 &  \nodata &  \nodata  & 11 $\pm$ 4 \\
 6300 $[$O I$]$   &$-$0.30& 5.5 $\pm$ 0.8 & 3.7 $\pm$ 1.1 &  5.0 $\pm$ 1.0 &	\nodata & \nodata	& \nodata\\
 6548 $[$N II$]$  &$-$0.34& 29.4 $\pm$ 1.5 &27.1 $\pm$ 2.4 & 26.9 $\pm$ 2.9 &   23 $\pm$  9  &   26 $\pm$ 6 & 17.6 %%@
$\pm$ 4.9 \\ 
 6563 H$\alpha$   &$-$0.34& 286 $\pm$ 10 &286 $\pm$ 14   &  286 $\pm$ 15  &  286 $\pm$ 83  &  286 $\pm$ 78  &  286 %%@
$\pm$ 37 \\
 6584 $[$N II$]$  &$-$0.34& 79.3 $\pm$ 3.1 & 78.9 $\pm$ 5.6 & 76.4 $\pm$ 5.5 &   62 $\pm$ 19  &   81 $\pm$ 19  &   %%@
58.2 $\pm$ 9.3 \\ 
 6678 He I	      &$-$0.35& 2.91 $\pm$ 0.38 & 2.3:           &  3.4:           & 	 \nodata & \nodata	& \nodata \\
 \noalign{\smallskip}
\hline
 \noalign{\smallskip}
 $F$(\Hb)$^b$&\nodata& 192 $\pm$ 6 &36.5 $\pm$ 1.3&28.7 $\pm$ 1.0&0.329 $\pm$ 0.097&0.324 $\pm$ 0.081& 2.39 $\pm$ %%@
0.31\\
 C(\Hb)      &\nodata& 0.22 $\pm$ 0.02 &0.30 $\pm$ 0.02&0.29 $\pm$ 0.02&0.18 $\pm$ 0.03&0.25 $\pm$ 0.03&0.18 $\pm$ %%@
0.03\\
 $W_{abs}$$\rm^c$   &\nodata& 1.8 $\pm$ 0.1 & 1.1 $\pm$ 0.1  &0.9 $\pm$ 0.1  & 1.5 $\pm$ 0.1 &1.5 $\pm$ 0.1  & 1.5 %%@
$\pm$ 0.1\\ 
 $-W$(\Ha)$\rm^c$   &\nodata& 104 $\pm$ 2 & 140 $\pm$ 6    &93 $\pm$ 4     & 18  $\pm$ 2   &26 $\pm$ 3     &  27 %%@
$\pm$ 0.02\\ 
 $-W$(\Hb)$\rm^c$   &\nodata& 19.0 $\pm$ 0.8 & 25  $\pm$ 2    &17.6 $\pm$ 1.5 & 5.6  $\pm$ 1.1&6.7 $\pm$ 0.7  & 7.5 %%@
$\pm$ 0.2\\
 $-W$(H$\gamma$)$\rm^c$    &\nodata& 5.9 $\pm$ 0.3 & 4.9 $\pm$ 0.4  &5.7 $\pm$ 0.6  &\nodata        &\nodata        %%@
&\nodata \\
 $-W$(H$\delta$)$\rm^c$    &\nodata& 2.1 $\pm$ 0.4 &1.2 $\pm$ 0.2  &2.3 $\pm$ 0.4  &\nodata        &\nodata        %%@
&\nodata\\
 $-W$([O III])$\rm^c$     &\nodata& 27.2 $\pm$ 0.9 & 24 $\pm$ 1   & 29 $\pm$ 1    & 4.0 $\pm$ 0.4 & 6.7 $\pm$ 0.4 & %%@
7.9 $\pm$ 0.4\\    
 \noalign{\smallskip}
    \hline\hline
  \end{tabular}
   \begin{flushleft}
   $\rm^a$ $[$O II$]$ $\lambda\lambda$3726,3729 emission flux.\\
   $\rm^b$ In units of 10$\rm^{-15}$ erg s$\rm^{-1}$ cm$\rm^{-2}$. \\
   $\rm^c$ In units of \AA.\\
   \end{flushleft}
\end{table*}

\subsection{Intermediate resolution spectra}

The slit position was placed crossing the center of both galaxies, with a position angle of 138$^\circ$. In %%@
Figure~\ref{figHa} we show the slit position over the \Ha\ map of IRAS 08339+6517; the \Ha\ flux distribution along %%@
the slit is shown in Figure~\ref{rendija}. We have also marked in this figure the apertures used to extract the %%@
individual 1-D spectra. The two brightest peaks correspond with the central object (knot \#1) and knot \#2 that we %%@
previously detected in the \Ha\ image. We have also extracted two external zones, A and B, where weak peaks of \Ha\ %%@
emission are detected. All four apertures have a size of 1$\arcsec \times 1\arcsec$. But we have also extracted four %%@
additional spectra with increasing apertures to analyze the dependence of the properties of the ionized gas with the %%@
aperture size. These new apertures have a size of 2.5$\arcsec$, 5$\arcsec$, 9$\arcsec$, and 12$\arcsec \times %%@
1\arcsec$, and are also indicated in Figure~\ref{rendija}. The best S/N ratio is found in the largest aperture, %%@
corresponding to the integrated spectrum of IRAS 08339+6517. This aperture collects approximately 96\% of the total %%@
flux of the galaxy observed by our slit. Only one spectrum of the companion galaxy was extracted. Its aperture size %%@
was 6.5$\arcsec\times 1\arcsec$.

The wavelength- and flux-calibrated spectra of the 12$\arcsec\times 1\arcsec$ aperture for IRAS 08339+6517 and the %%@
single aperture for the companion galaxy are shown in Figure~\ref{spectra}. The spectrum of IRAS 08339+6517 shows a %%@
high conti\-nuum level and an intense underlying stellar absorption in \ion{H}{i} Balmer lines. All of them, except %%@
\Ha, show intense absorption wings. This was previously noted by \citet{MHK03}. Furthermore, we have detected the %%@
main absorption features present in this spectral range: \ion{Ca}{ii} H $\lambda$3933, G-band $\lambda$4304, %%@
\ion{Mg}{i} $\lambda\lambda$5167,87 and Na D1 $\lambda$5889, as well as some \ion{He}{i} absorption. These features, %%@
together the high continuum level of the spectrum, can be interpreted as the product of a substantial population of %%@
older stars, with ages much greater than 10 Myr. Consequently, at least two separate stellar populations are present %%@
in the galaxy (see \S4.1). The spectrum of the companion object is noisier but shows a rather flat low continuum %%@
level with no important absorption features, characteristic of an \ion{H}{ii} region-dominated zone.

Fluxes and equivalent widths of the emission lines were measured using the SPLOT routine of the IRAF package. This %%@
task integrates all the flux in the line between two given limits and over a fitted local continuum. All the emission %%@
lines detected were corrected for reddening using the \citet{seaton79} law and the corresponding value of C(H$\beta$) %%@
derived in the way we explained in \S3.1. The H$\beta$ line flux, $F(H\beta)$ and the equivalent width of several %%@
lines (H$\alpha$, H$\beta$, H$\gamma$, H$\delta$ and [\ion{O}{iii}]) are given in Table~\ref{table4}, as well as the %%@
reddening-corrected line intensity ratios relative to H$\beta$ and their uncertainties. Colons indicate uncertainties %%@
greater than 40\%. Note that the values obtained for the \Ha\ equivalent width of \#1, \#2, IRAS 08339+6517 and the %%@
companion galaxy using the spectra and the \Ha\ image are similar to within the uncertainties.
% This indicates the goodness of our reduction and analysis procedures for both images and spectra. 

\begin{table*}[th]\centering
 \caption{Chemical abundances of the galaxies and analyzed knots inside IRAS 08339+6517.}
 \label{table5}
 \footnotesize
 \begin{tabular}{lcccccc}
 \noalign{\smallskip}
   \hline\hline
   \noalign{\smallskip}
Object    & IRAS 08339+6517  & \#1     &   \#2         & A       &  B           &   Companion \\
    \hline
    \noalign{\smallskip}
$T_e[$O III$]\rm^a$    & 8700 & 7900   & 7900          & 8700    &  7700          & 9050  \\
$T_e[$O II$]\rm^b$     & 9100 & 8500   & 8500          & 9100    &  8400          & 9300  \\
%    \noalign{\smallskip
12+log O/H$^c$          &8.45 $\pm$ 0.10$\rm^f$& 8.55 $\pm$ 0.10&8.53 $\pm$ 0.10&8.42 $\pm$ 0.10&8.58 $\pm$ 0.10 & %%@
8.38 $\pm$ 0.10 \\
12+log O/H$^d$          &8.71 $\pm$ 0.20& 8.71 $\pm$ 0.20&8.70 $\pm$ 0.20&8.63 $\pm$ 0.20&8.72 $\pm$ 0.20 & 8.62 %%@
$\pm$ 0.20  \\
(O$^{++}$+O$^+$)/O$^+$  &1.53     &  1.85     & 1.63         & 1.29        &  2.04         & 1.81    \\
log N$^+$/O$^+$         & $-$0.94 & $-$0.89   & $-$0.94      & $-$1.10     & $-$0.86       & $-$1.13   \\
12+log N/H              & 7.51    &  7.65     & 7.59         & 7.21        & 7.72          & 7.26   \\
12+log Ne$^{++}$/H$^+$  & 7.54    &  7.73     & 7.56         &\nodata      & \nodata       & 7.48:   \\
12+log He$^+$/H$^+$     & 10.91 $\pm$ 0.03    &  10.98 $\pm$ 0.06  & 10.92 $\pm$ 0.07  &\nodata      & \nodata       %%@
& 10.90 $\pm$ 0.13 \\
12+log Fe$^{++}$/H$^+$ &  5.95 $\pm$ 0.16    &  5.91:   & \nodata   &\nodata      & \nodata       & \nodata \\
 \hline
 \noalign{\smallskip}
[O/H]/[O/H]$_{\odot}$$^e$&0.62    & 0.78         & 0.74          & 0.58           & 0.83          & 0.52 \\
 \hline
 \hline
 \end{tabular}
 \begin{flushleft}
  $\rm^a$ In units of $10^4$K, calculated using empirical calibrations of \citet{P01}\\
 $\rm^b$ In units of $10^4$K, $T_e [$OII$]$ calculated from empirical calibration and the relation given by %%@
\citet{G92}\\
 $\rm^c$ Determined using the empirical calibration of \citet{P01}\\
 $\rm^d$ Determined using the $[$N II$]$/H$\alpha$ ratio \citep{D02}\\
 $\rm^e$ Asuming 12+$\log$[O/H]$_\odot$ = 8.66$\pm0.05$ \citep{ASP04}\\
 $\rm^f$ The oxygen abundance derived for this object using the direct method is 8.42\\
 \end{flushleft}
\end{table*}

\subsubsection{Physical conditions of the ionized gas}

We have studied the physical conditions and chemical abundances of the ioni\-zed gas from our 1-D spectra. Although %%@
the weak [\ion{O}{iii}] $\lambda$4363 emission line is barely detected in the brightest galaxy, the underlying %%@
stellar absorption affecting $H\gamma$ does not permit a proper measurement of that auroral line, except for the %%@
12$\arcsec$ aperture (see below). The oxygen abundance was derived then using the \citet{P01} empirical calibration. %%@
The electron temperatures have been estimated from the T([\ion{O}{iii}]) and T([\ion{O}{ii}]) pairs that reproduce %%@
the total oxygen abundance obtained by applying the \citet{P01} empirical method (see next section). Like Pilyugin, %%@
we have assumed a two-zone scheme, the usual relation O/H = O$^+$/H$^+$ + O$^{++}$/H$^+$ and the linear relation %%@
between T([\ion{O}{iii}]) and T([\ion{O}{ii}]) based on photoionization models obtained by \citet{G92}. The finally %%@
adopted T[\ion{O}{iii}] and T[\ion{O}{ii}] for each burst are shown in Table~\ref{table5}. We have no determination %%@
of electron densities, so we assumed that $N_e$ = 100 cm$^{-3}$ for all the objects. 
%buena aproximacion, si ne=500, las abundancias serian +0.02 dex).

The spectrum obtained using the 12$\arcsec$ aperture allows a tentative determination of the electron temperature of %%@
the ionized gas because of the detection of the [\ion{O}{iii}] $\lambda$4363 and [\ion{N}{ii}] $\lambda$5755 emission %%@
lines. These weak auroral emission lines have an error of $\sim$54\% and 34\%, respectively, giving T([\ion{O}{iii}]) %%@
= 8900$^{+1200}_{-1600}$ and T([\ion{N}{ii}]) = 9000$^{+2600}_{-1300}$. These values are in agreement with those we %%@
obtained using the empirical calibration. Furthermore, the direct determination also agrees with the \citet{G92} %%@
linear relation between the high and low electron temperatures.

%\begin{eqnarray}
%T([\textsc{O~ii}]) = 0.7 \times T([\textsc{O~iii}]) + 3000. 
%\end{eqnarray}

\subsubsection{Abundance Analysis}

\citet{P01} performed a detailed analysis of observational data combined with photoionization models to obtain %%@
empirical calibrations for the oxygen abundance from the relative intensities of strong optical lines. He used both %%@
the $R_{23}$ and $P$ (an indicator of the hardness of the ionizing radiation) parameters for his calibration, the %%@
so-called \emph{P-method}, which can be used in moderately high-metallicity \ion{H}{ii} regions (12+$log$ O/H $\geq$ %%@
8.15). We have used this empirical calibration to derive the total oxygen abundance in IRAS 08339+6517 and its %%@
companion galaxy. We have also estimated the total oxygen abundance making use of the \citet{D02} empirical %%@
calibration, which involves the [\ion{N}{ii}] $\lambda$6583 /\Ha\ ratio. The oxygen abundances obtained using both %%@
methods are shown in Table~\ref{table5}. However, the O/H ratios obtained using the last calibration seem to be %%@
systematically higher than those obtained using the \citet{P01} method, as we reported in a previous study %%@
\citep{LSER04b}. Here, an offset of 0.15--0.20 dex is found between both calibrations. We prefer to consider the %%@
\citet{P01} abundances as the more appropriate ones because his calibration take into account the excitation %%@
parameter of the ionizing radiation. 
%Furthermore, the uncertainty that he gives for his calibration is $\pm 0.10$ dex, 
%whereas \citet{D02} estimate an error of $\sim\, \pm 0.2$ dex for their oxygen abundances. 

All the ionic abundances shown in Table~\ref{table5} (except He$^+$ and Fe$^{++}$) have been calculated using the %%@
IRAF NEBULAR task from the intensity of collisionally excited lines, assuming the T([\ion{O}{iii}]) for the high %%@
ionization potential ions O$^{++}$ and Ne$^{++}$, and T([\ion{O}{ii}]) for the low ionization potential ions O$^+$ %%@
and N$^+$. We have assumed the standard ionization correction factor (icf) by \citet{PC69}, N/O = N$^+$/O$^+$, to %%@
derive the nitrogen abundance of the objects. The He$^+$/H$^+$ ratio has been derived from the \ion{He}{i} %%@
$\lambda$5876 line observed in each knot, using the predicted line emissivities calculated by \citet{SSM96}. We have %%@
also corrected for collisional contribution following the calculations by \citet{B02}. Self-absorption effects were %%@
not considered for the determination of He$^+$/H$^+$ but only errors derived from the observed uncertainties in line %%@
measurement. 

We have detected the [\ion{Fe}{iii}] $\lambda$4658 emission line in the integrated spectrum of the main galaxy and in %%@
knot \#1. The calculations for Fe$^{++}$ have been done with a 34 level model-atom that uses the collision strengths %%@
of \citet{Z96} and the transition probabilities of \citet{Q96}. Applying the relation given by \citet{RR05}, we %%@
derived a total Fe abundance of 12+log(Fe/H)=6.11 and log(Fe/O)=$-$2.60 for the galaxy.

In general, we find that the O/H ratio of the knots inside IRAS 08339+6517 are similar [around 8.5, in units of 12 + %%@
log(O/H)], except object A, which has a slightly lower value [12+log(O/H)=8.42]. 
%It is interesting to remark that A and B have very different oxygen abundances despite they are located at similar %%@
%distance from the centre of the galaxy (see Figure~\ref{rendija}). We will analyze it in \S4.4. 
On the other hand, the oxygen abundance of the companion galaxy is a bit lower than that derived for IRAS 08339+6517. %%@
The N$^+$/O$^+$ ratios are similar for all the objects, within the expected values for spiral and irregular galaxies %%@
with the same oxygen abundance \citep{G99}. These facts suggest that both galaxies have suffered a comparable degree %%@
of chemical enrichment despite their different absolute magnitudes. 

\subsubsection{Kinematics of IRAS 08339+6517}

\begin{figure}[t]
\centering
\includegraphics[angle=270,width=1\linewidth]{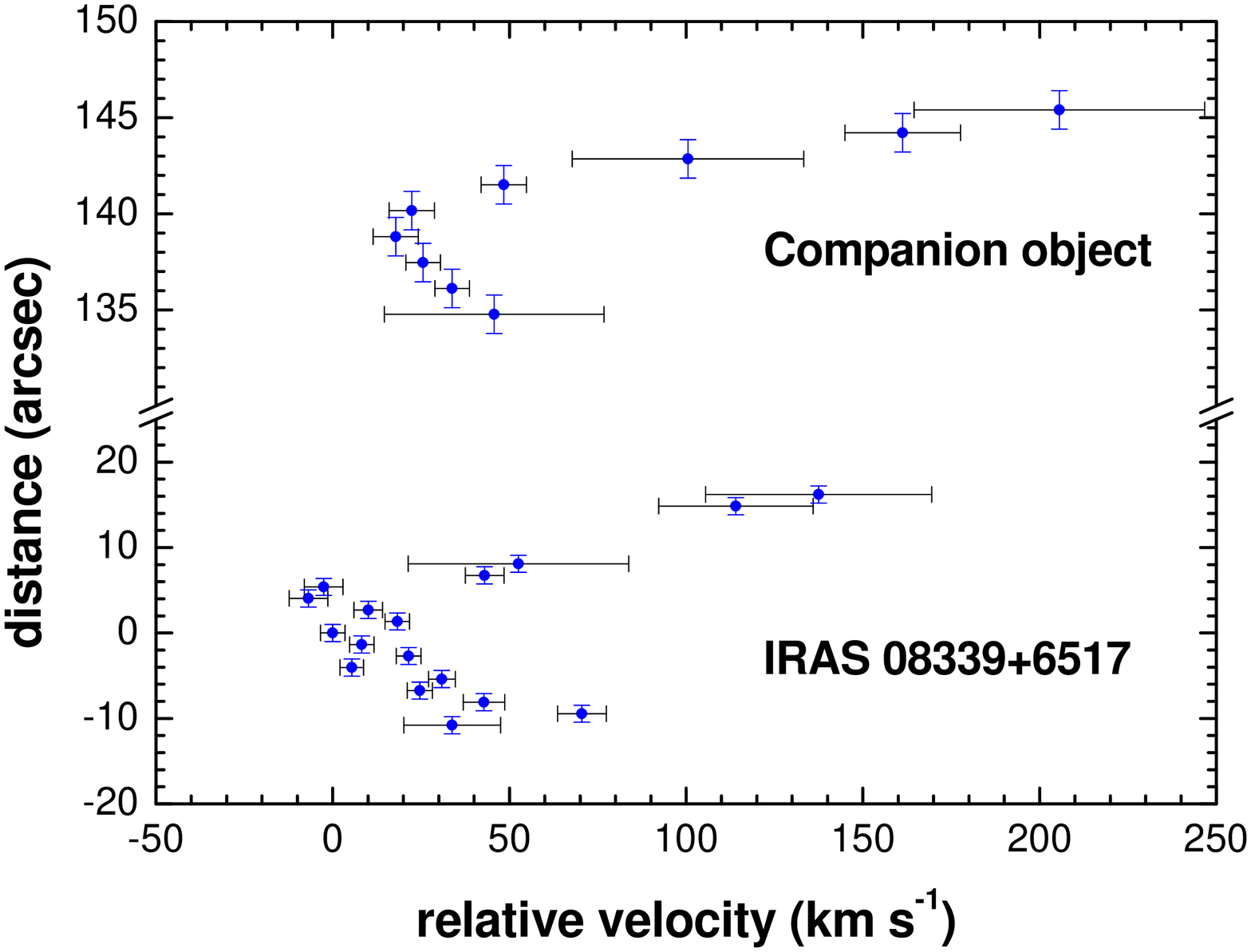}
\protect\caption[ ]{\small{Position--velocity diagram for the slit position observed, analyzed in 1.1$\arcsec$ bins. %%@
The horizontal bars represent the uncertainty of the Gaussian fitting for each point. Note that the vertical axis has %%@
been broken for clarity. Northwest is up.}}
\label{pv}
\end{figure}

The kinematics of the ionized gas were studied via the spatially resolved analysis of the \Ha\ line profile along the %%@
slit position. We have extracted zones 6 pixels wide (1.1$\arcsec$ long) covering all the extension of the %%@
line-emission zones. The Starlink DIPSO software \citep{HM90} was used to perform a Gaussian fit to the H$\alpha$ %%@
profiles. For each single or multiple Gaussian fit, DIPSO gives the fit parameters (radial velocity centroid, %%@
Gaussian sigma, FWHM, etc) and their associated statistical errors. In Figure~\ref{pv} we show the position-velocity %%@
diagram obtained, indicating the position of the studied objects. All the velocities are referred to the heliocentric %%@
velocity of the center of IRAS 08339+6517 (5775 km s$^{-1}$).
%(5960 km s$^{-1}$, value that was not corrected by the heliocentric velocity for the rotational motion the Sun about %%@
%the galaxy nor the infall of the Local Group toward the Virgo Cluster).
 
The main galaxy has an apparent global rotation pattern, but it also shows local velocity fluctuations that could be %%@
produced by the effects of the very recent star formation activity (gas motions due to the combination of the winds %%@
of massive stars and supernova explosions) or by distortions in the gas associated with interaction effects (with the %%@
companion galaxy or, perhaps, with an hypothetical pre-existing external object that could have merged with IRAS %%@
08339+6517 in the past). Interaction features are clearly detected in the outer northwest area of IRAS 08339+6517, %%@
where the ionized gas does not follow the general kinematics but presents a continuous variation between +45 and +130 %%@
km s$^{-1}$. These positions coincide with the \Ha\ emission detected in knot B and the NW arm. The companion galaxy %%@
seems to shows an apparent rotation behavior in its southeast area (its brighter zone), although an important %%@
velocity gradient (around +200 km s$^{-1}$) is found in the side opposites the main galaxy. This fact suggests again %%@
that gravitational interactions have affected the dynamics of the ionized gas.

\begin{figure}[t!]
\includegraphics[angle=270,width=1\linewidth]{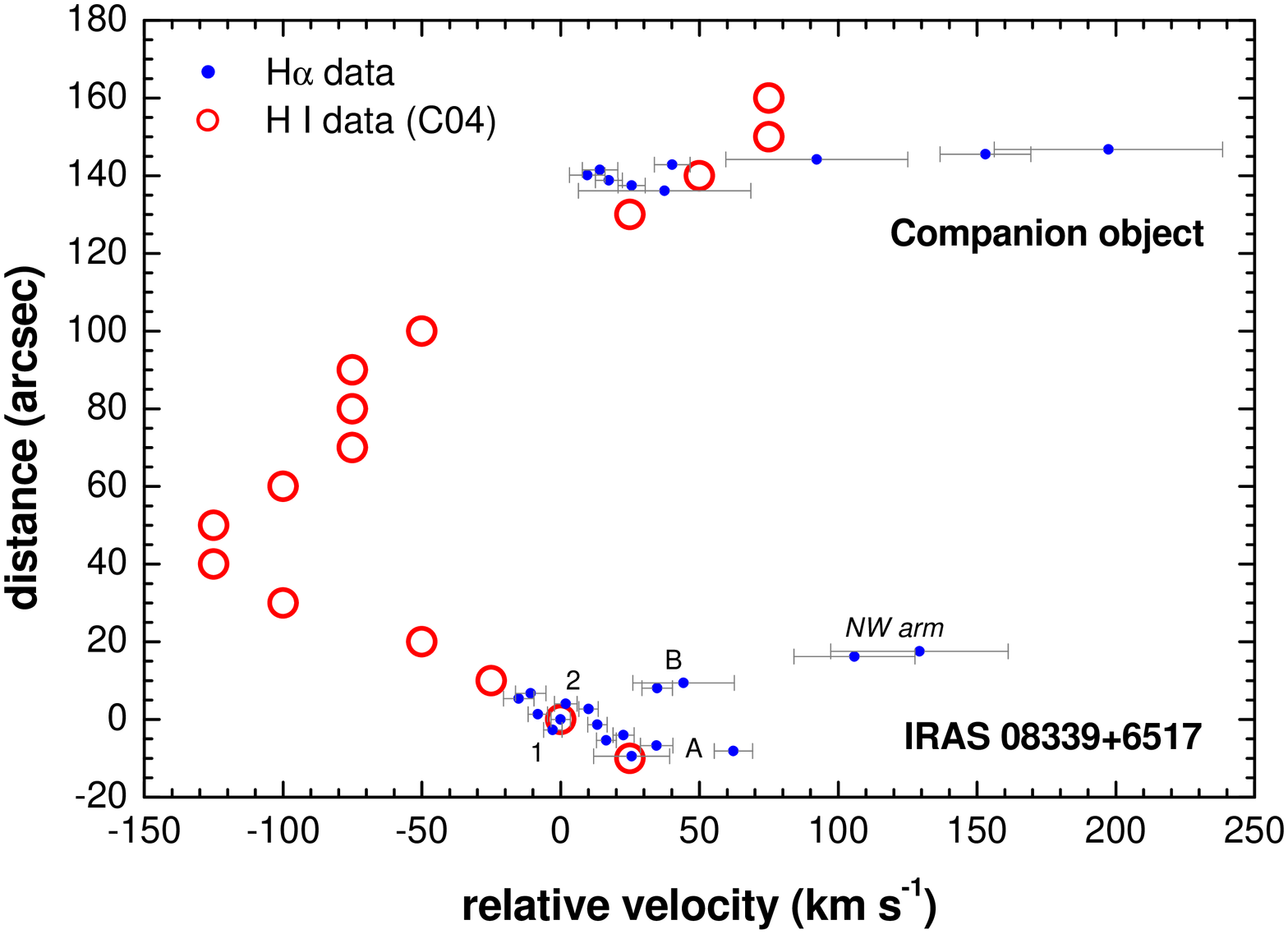}
\caption{\small{Position--velocity diagram shown in Figure~\ref{pv} compared with that obtained using the \ion{H}{i} %%@
data from \citet{CSK04} (open circles). The outflow of the neutral gas is clearly observed, as well as the decoupling %%@
of the ionized gas at the NW arm of IRAS 08339+6517.}}
\label{pv2}
\end{figure}

Assuming the global kinematics of the galaxies as circular rotation, we can make a rough estimate of their Keplerian %%@
mass, $M_{kep}$, taking into account the half of the maximum velocity difference in the galaxy and the half of the %%@
spatial separation corresponding to these maxima. We do not consider in this calculus the effect of the velocity %%@
gradients presented in both galaxies. We obtain masses of (10$\pm$3)$\times$10$^9$ \Mo\ and (8$\pm$2)$\times$10$^8$ %%@
\Mo\ for IRAS 08339+6517 and its companion galaxy, respectively, assuming circular orbits, Keplerian dynamics and an %%@
inclination of 90$^\circ$. We remark again that these values are only very tentative estimations of the masses of the %%@
galaxies. 

We have compared the Keplerian mass, the neutral gas mass, $M_{H\,I}$, and the ionized hydrogen mass, $M_{H\,II}$ %%@
(see Table~\ref{table2}) of the both galaxies and find substantial differences between them. IRAS 08339+6517 has %%@
higher $M_{H\,II}$/$M_{H\,I}$ and $M_{kep}$/$M_{H\,I}$ ratios (0.013 and 9.1, respectively), while the companion %%@
object has 0.0004 and 1.1.
%suggesting that it has transformed more efficiently its neutral gas to ionized gas and stars.
However, if we consider that the tidal \ion{H}{i} gas found between both objects \citep{CSK04}, with a mass of %%@
$M_{H\,I}$ = (3.8$\pm$0.5)$\times$10$^9$ \Mo, have been expelled from the main galaxy, the new $M_{kep}$/$M_{H\,I}$ %%@
ratio of IRAS 08339+6517 is 2.0, similar to that found in the companion galaxy. 

The light-to-mass ratios, $L_B/M_{\rm{kep}}$ and $L_B/M_{\rm{H\,I}}$, are also very different for both galaxies. The %%@
main object has $L_B/M_{\rm{H\,I}} \sim$ 61, much higher than the typical values for spiral and irregular galaxies, %%@
that are between 1.6 and 11.2 \citep{BGGB03}. The companion galaxy has a value between those limits %%@
($L_B/M_{\rm{H\,I}} \sim$ 4.2), similar to an Sbc galaxy. If we again consider that the tidal neutral gas belongs to %%@
IRAS 08339+6517, the new light to neutral gas ratio is $\sim$13.7, only a bit higher than the expected value for an %%@
Sa galaxy (11.2 $\pm$ 2.0). The light-to-cold dust ratio, L$_B$/M$_{dust}\ \sim\ 1.5\times 10^4$, is also the %%@
expected one for young spirals \citep{BGGB03}. The mass of the dust was calculated using the FIR 60$\mu$m and %%@
100$\mu$m flux densities \citep{Mo90} together with the relation given by \citet{HSH95} (their last equation in p. %%@
678). All these facts support the idea that the tidal \ion{H}{i} gas has been formed from material mainly stripped %%@
from IRAS 08339+6517. 

%\begin{eqnarray}
%\frac{M}{M_\odot} \sim 233 \times d~[\rm{pc}] \cdot 
%\bigg(\frac{v~[\rm{km~s^{-1}}]}{\sin i}\bigg)^2
%\end{eqnarray}

In order to improve our knowledge of the kinematics of the system, we have compared the position--velocity dia\-gram %%@
obtained from our analysis of the \Ha\ line profile in the optical spectrum with the position--velocity diagram %%@
obtained from \ion{H}{I} data. We have used the Figure~2 of \citet{CSK04} to extract the velocity of the neutral gas %%@
along our slit position (with P.A. 138$^{\circ}$) in 10$\arcsec$ steps. We referred all the \ion{H}{i} velocities to %%@
the heliocentric velocity of the center of IRAS 08339+6517 (5775 km s$^{-1}$). In Figure~\ref{pv2} we show the %%@
position--velocity diagram combining \ion{H}{i} and \Ha\ data. 

The general correspondence between both diagrams is rather good: the centers of the two galaxies and the kinematics %%@
of IRAS 08339+6517 practically coincide in both neutral and ionized gas. \citet{CSK04} have remarked that, although %%@
the companion galaxy shows a component of solid-body rotation in neutral gas, the signs of rotation are less %%@
prominent in the main galaxy. But it could be masked in part by the outflow of the neutral gas in the direction of %%@
the observer, as is clearly observed in Figure~\ref{pv2}. This outflow of neutral gas was firstly suggested by %%@
\citet{GDL98} because of the excess absorption detected in the blue wing of the \ion{O}{vi} $\lambda$1032 profile. %%@
They also noted that the broadening of the Ly$\alpha$, \ion{Si}{ii} $\lambda$1260 and \ion{C}{ii} $\lambda$1335 %%@
absorption lines (the metallic lines are also resolved in two components) indicated large-scale motions of the %%@
interstellar gas, with a velocity even higher than 3000 km s$^{-1}$. \citet{GDL98} suggested that the kinematics of %%@
the warm ionized interstellar medium are not driven by their gravitational potential but by the dynamical %%@
consequences of the violent star formation processes taking place in the galaxy. \citet{KMH98} attributed the %%@
existence of a secondary Ly$\alpha$ emission peak to the chaotic structure of the interstellar medium. Our \Ha\ data %%@
suggest that, although these significant distorsions are found in the ionized gas, the kinematics of the galaxy still %%@
have an apparent important component of solid-body rotation.    

However, we find a clear divergence between the kinematics of the neutral and the ionized gas at the NW arm of IRAS %%@
08339+6517, where object B is located: while the first has a velocity of $-$50 km s$^{-1}$ with respect to the center %%@
of the galaxy, the second shows a velocity around +100 km s$^{-1}$. Thus, the ionized gas seems to be decoupled from %%@
the expanding motion of the tidal tail of neutral gas. This suggests that the interaction that has created the %%@
\ion{H}{I} tail is more complex that we thought. Knot B is specially interesting because it has the hightest oxygen %%@
abundance (12 + log O/H = 8.58), being the higher of all the spectroscopically observed regions (see %%@
Table~\ref{table5}).

\section{Discussion}

\subsection{Ages of the bursts and stellar populations}

\subsubsection{Ages from $W$(H$\alpha$)}

One of the best methods of determining the age of the recent bursts of star formation is through the \Ha\ equivalent %%@
width, $W$(H$\alpha$), since it decreases with time. We have used the STARBURST99 \citep{L99} spectral synthesis %%@
models to estimate the age of each knot, comparing the predicted \Ha\ equivalent widths with the observed values. %%@
Spectral synthesis models with two different metallicities were chosen (Z/Z$_\odot$ = 1 and 0.4, the appropriate %%@
range of metallicities for the objects, see Table~\ref{table5}), both assuming an instantaneous burst with a Salpeter %%@
IMF, a total mass of 10$^6$ $M_\odot$ and a 100 $M_\odot$ upper stellar mass. The ages obtained are compiled in %%@
Table~\ref{table3}. All of them show a very young age for the last star formation episode, between 3.5 and 6 Myr, %%@
being the center of the galaxy (knot \#1) the youngest one. We can adopt an average age of 4 -- 5 Myr for the most %%@
recent star formation episode in IRAS 08339+6517 and 5 -- 6 Myr in its companion galaxy. 

\begin{figure}[t!]
\includegraphics[angle=90,width=1\linewidth]{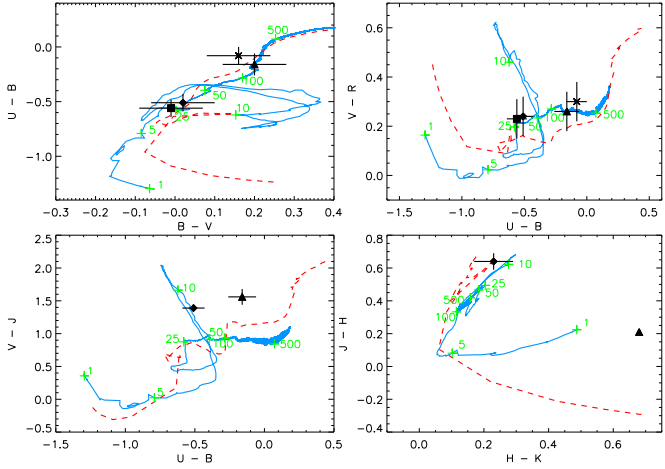}
\caption{\small{Color-color diagrams showing the STARBURST99 \citep{L99} (blue/continuous line) and PEGASE.2 %%@
\citep{PEGASE97} (dashed/red line) models for an instantaneous burst with a Salpeter IMF and $Z/Z_{\odot}$ = 0.4 %%@
compared with our observed values for IRAS 08339+6517 (circle) and its companion galaxy (triangle). We have also %%@
included the derived values for the underlying component in IRAS 08339+6517 (cross) and the burst component (square). %%@
We indicate some age ticks (in Myr) along the plotted evolutionary tracks for the STARBURST99 model.}}
\label{models}
\end{figure}

\subsubsection{Ages from evolutionary synthesis models}

\citet{GDL98} used evolutionary synthesis models to constrain the star formation history of IRAS 08339+6517 from  %%@
their UV spectra. They found that a 6--7 Myr burst with $M_{up} \geq 30$ \Mo\ or 9 Myr continuous star formation with %%@
$M_{up} \leq 30-40$ \Mo\ are compatible with their observations. They noted that the weakness of the stellar UV %%@
absorption lines and the small \Ha\ equivalent width suggested significant dilution by an underlying stellar %%@
population. Our estimated ages from $W$(H$\alpha$) are a little lower than those they predicted. These authors also %%@
compared the \ion{C}{iv} and \ion{Si}{iv} profiles with those of the starburst galaxy NGC 1741, suggesting that IRAS %%@
08339+6517 is probably in a more advanced evolutionary stage. But the weakness of these lines could also be explained %%@
if the IMF is steeper or there exists a dilution of the stellar lines due to the underlying population. The latest %%@
study of NGC~1741 (member C in HCG~31) performed by \citet{LSER04a} estimates an age of around 5 Myr and %%@
$E(B-V)=$0.06 for this galaxy. Thus, IRAS 08339+6517 and NGC 1741 have very similar ages, but the first has a larger %%@
reddening contribution. Perhaps the effects of an underlying population is the most important factor in the dilution %%@
of the UV stellar lines in the \citet{GDL98} study.

We have compared our reddening-corrected colors with STARBURST99 \citep{L99} and PEGASE.2 \citep{PEGASE97} models to %%@
get an additional estimation of the age of the two galaxies. We have chosen these two models because while the firsts %%@
are based on Geneva tracks, the seconds use Padua isochrones \citep{padua94} in which thermally pulsing asymptotic %%@
giant branch (TP-AGB) phases are included. We assumed an instantaneous burst with a Salpeter IMF, a total mass of %%@
$10^6$ \Mo\ and a metallicity of $Z/Z_{\odot}$ = 0.4 and 1 for both models. In Figure~\ref{models} we plot four %%@
different color-color diagrams comparing both STARBURST99  and PEGASE.2 $Z/Z_{\odot}$ = 0.4 models with the observed %%@
values. We find a good correspondence using our optical values although PEGASE.2 give older ages than STARBURST99. In %%@
particular, the $V-R$ color of IRAS 08339+6517 implies an age of 40 -- 60 Myr using STARBURST99 models, but 450 -- %%@
600 Myr when it is compared with PEGASE.2 models. Furthermore, the NIR colors derived from 2MASS %%@
(reddening-corrected, see Table~\ref{table2}) are not in agreement with optical colors, specially in the case of the %%@
companion galaxy. The optical colors suggest an age of $\sim$ 30 -- 50 Myr for IRAS 08339+6517 and older than 200 Myr %%@
for its companion. The NIR colors imply ages older than 1 -- 2 Gyr for both systems. The considerable difference in %%@
the age derived for the companion galaxy using the $W$(H$\alpha$) and the optical and NIR colors indicates that the %%@
old populations practically dominate the observed flux at these long wavelengths.  

\begin{figure}[t!]
\includegraphics[angle=270,width=1\linewidth]{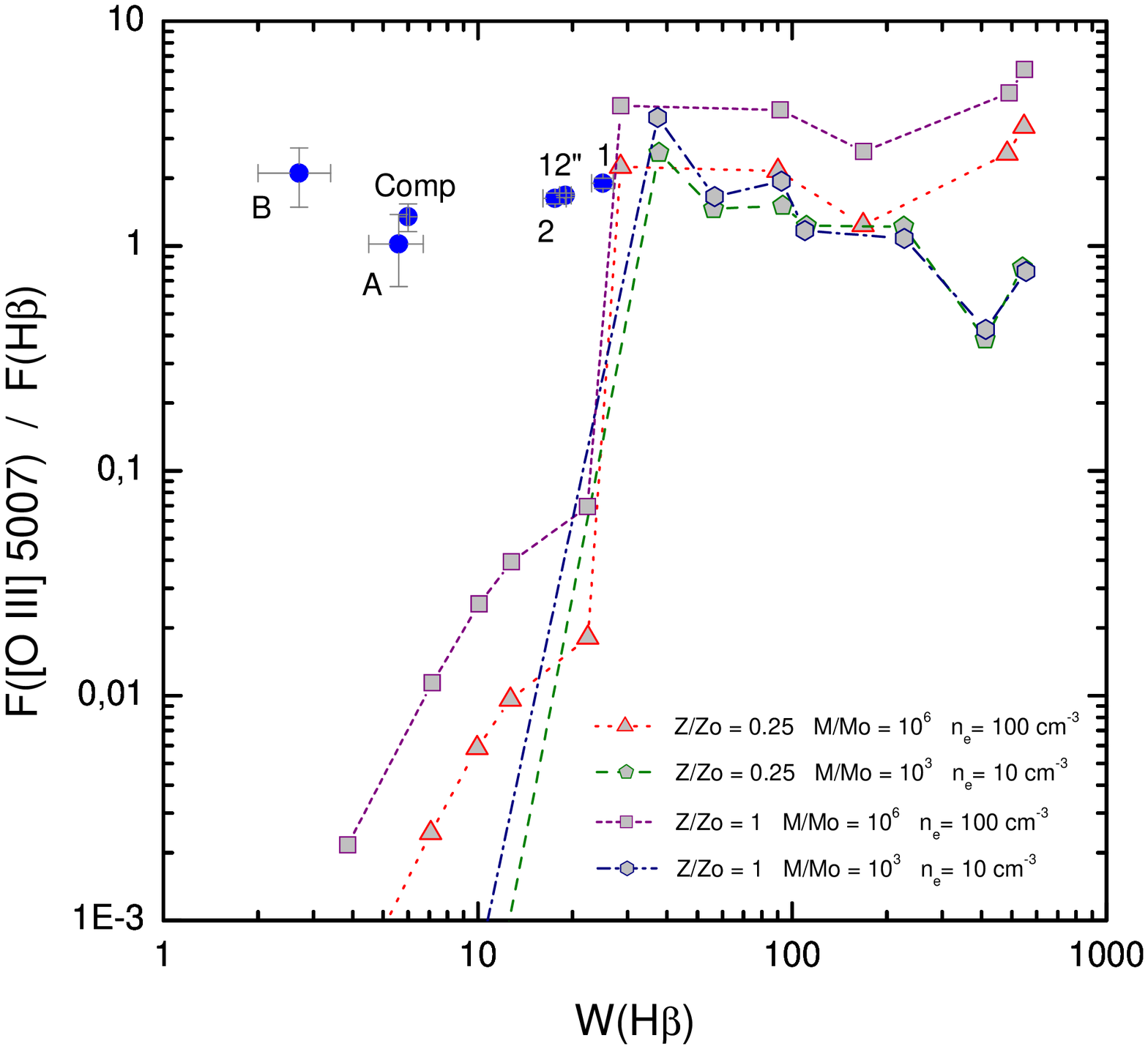}
\caption{\small{$F[\ion{O}{iii}]\ \lambda 5007$ versus $W$(\Hb) using the models by \citet{SSL01}. Tracks correspond %%@
to sequences of different metallicities and electronic densities. Each symbol marks the position of the
models at 1 Myr interval, starting in the upper-right corner of the diagram with an age of 1 Myr.}}
\label{sta}
\end{figure}

The fact that the ages derived from colors are older than those obtained from $W$(H$\alpha$) confirms that the %%@
effects of the underlying old population are important. The contribution of the old stellar population can be checked %%@
using the \citet{SSL01} models of \ion{H}{ii} regions ionized by an evolving starburst embedded in a gas cloud of the %%@
same metallicity. We have chosen two models, both with metallicity Z/\Zo\ = 0.25 and 1 and with a total mass of %%@
10$^3$ and 10$^6$ \Mo. In Figure~\ref{sta} we plot our observational values of $W$(\Hb) and the [\ion{O}{iii}] %%@
$\lambda$5007/\Hb\ emission line flux (see Table~\ref{table3}) and compare them with the theoretical models. We find %%@
that all are located in the left area of the diagram, indicating a strong contribution of the underlying stellar %%@
continuum. Objects \#1 and \#2 show the best correspondence because they are regions mainly dominated by the recent %%@
star formation activity. The age derived for the objects following Figure~\ref{sta} is between 5 and 6 Myr, in %%@
agreement with those derived from $W$(H$\alpha$).  

\subsubsection{Analysis of the surface brightness profiles}

\begin{figure}[t!]
\includegraphics[angle=90,width=1\linewidth]{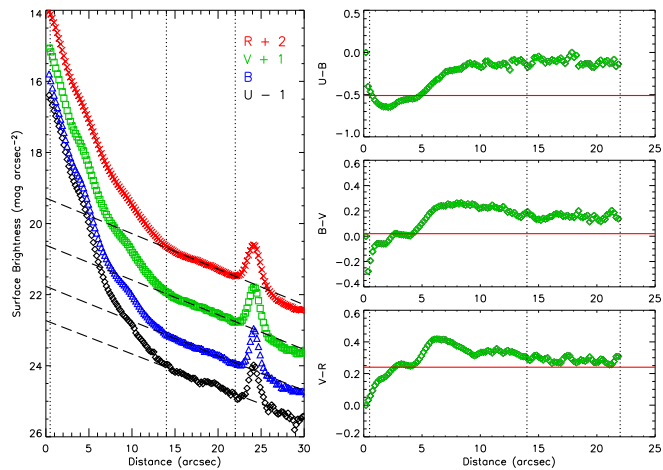}
\caption{\small{Surface brightness and color profiles for IRAS 08339+6517. The dashed lines in the surface brightness %%@
diagram represent exponential law fits to the profiles. The red horizontal lines in the color profile diagrams %%@
indicate the average color derived for each system. The dashed vertical lines indicate the radius of the seeing %%@
(0.4$\arcsec$) and the interval used to perform the fit (between 14$\arcsec$ and 22$\arcsec$). The peak at %%@
24$\arcsec$ is consequence of the bright star located at the SW of the galaxy.}}
\label{p1}
\end{figure}

\begin{table*}[t]\centering
 \caption{Structural parameters of IRAS 08339+6517 and its companion galaxy. The total luminosity in each band and %%@
the relative contribution of the underlying component at some radii are also shown. }
 \label{perfiles}
 \footnotesize
 \begin{tabular}{cccccccccccc}
 \noalign{\smallskip}
   \hline\hline
   \noalign{\smallskip}
& & &   IRAS 08339+6517 & & & & &   & Companion \\
Filter    & $\mu_0$  &   $\alpha$     &   $L_{total}$   &   \% UC      &   
\% UC    & \% UC     &   $\mu_0$  &   $\alpha$     &   $L_{total}$   &   \% UC\\
   &  (mag $\arcsec^{-2}$)     &  (kpc)  &  ($10^{43}$ erg s$^{-1}$) &  $r=2.5\arcsec$   &  $6\arcsec$ &  $14\arcsec$                            %%@
&  (mag $\arcsec^{-2}$) & (kpc)  &   ($10^{41}$ erg s$^{-1}$) &  $r=3.5\arcsec$\\ 
\hline
   \noalign{\smallskip}
$U\rm^a$&21.73& 4.49 &2.15$\pm$ 0.08        &  1 & 13 & 92 &   19.41 & 0.76 &  7.1 $\pm$ 0.5 & 84\\
$B$ & 21.77 & 4.31 & 3.09 $\pm$ 0.11        &  2 & 17 & 99 &   19.51 & 0.73 & 14.0 $\pm$ 0.6 & 92\\
$V$ & 21.61 & 4.31 & 1.72 $\pm$ 0.07        &  3 & 16 & 96 &   19.31 & 0.74 &  9.2 $\pm$ 0.5 & 95\\
$R$ & 21.28 & 4.21 & 2.20 $\pm$ 0.09$\rm^b$ &  3 & 14 & 96 &   19.26 & 0.83 & 11.9 $\pm$ 0.6 & 95\\
   \noalign{\smallskip}
 \hline
 \hline
 \end{tabular}
 \begin{flushleft}
  $\rm^a$ The errors in the fitting parameters in $U$ filter are higher than those found in the other filters. See %%@
text.\\ 
  $\rm^b$ Considering the \Ha\ emission (see Table~\ref{table3}), the total luminosity in $R$ filter is (2.08 $\pm$ %%@
0.10) $\times$ 10$^{43}$ erg s$^{-1}$. \\
 \end{flushleft}
\end{table*} 

\begin{figure}[t!]
\includegraphics[angle=90,width=1\linewidth]{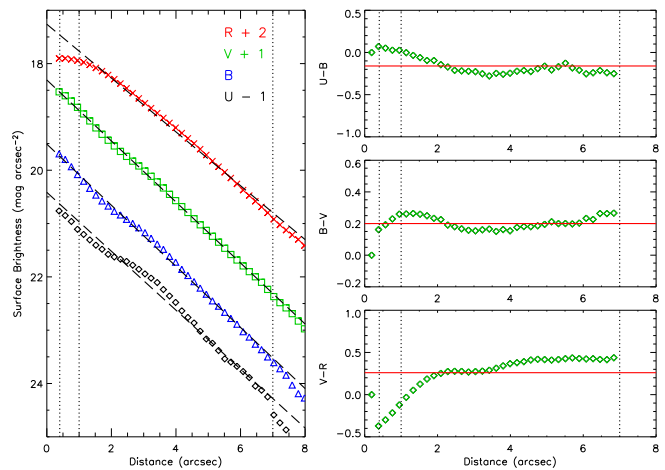}
\caption{\small{Surface brightness and color profiles for the companion galaxy. The dashed lines in the surface %%@
brightness diagram represent exponential law fits to the profiles. The red horizontal lines in the color profile %%@
diagrams indicate the average color derived for each system. The dashed vertical lines indicate the radius of the %%@
seeing (0.4$\arcsec$), and the interval used to perform the fit (between 2$\arcsec$ and 7$\arcsec$,
the apparent optical radius of the galaxy derived from the deep optical images). Note the bump at $\sim$3.5$\arcsec$ %%@
in $U$-filter, it is indicating the star-formation zone off-center this dwarf galaxy. }}
\label{p2}
\end{figure}

In order to investigate the importance of older stellar popu\-lations we have performed an analysis of the surface %%@
brightness profiles of the two galaxies. We have taken concentric surfaces at different radii from the center of each %%@
system and calculated the integrated flux inside each circle of area $A$ (in units of arcsec$^{2}$). Then, the mean %%@
surface brightness inside this circle, $SB_{X}$ (in units of mag arcsec$^{-2}$), is derived using the relation:
\begin{eqnarray}
SB_{X}=m_X+2.5 \log A ,
\end{eqnarray}
being $m_X$ the magnitude in the filter $X$. The surface brightness, $\mu_X$, is the flux per square arcsec in the %%@
ring defined by two successive apertures. This simple technique is not adequate for the study of objects presenting %%@
irregular or complex morphologies  \citep{CV01}, but it is valid for approximately circular compact objects like IRAS %%@
08339+6517 and its companion galaxy. In Figure~\ref{p1} we show the surface brightness for $U$, $B$, $V$ and $R$ %%@
filters, $\mu_U$, $\mu_B$, $\mu_V$ and $\mu_R$, versus its radius, for IRAS 08339+6517, whereas the surface %%@
brightness profiles for the companion galaxy are shown in Figure~\ref{p2}. In both Figures~\ref{p1} and \ref{p2} we %%@
also include the radial color profiles $(U-B)$, $(B-V)$ and $(V-R)$ derived by direct subtraction of light profiles.  

In the main galaxy, the surface brightness profiles can be separated into two structures, indicating the existence of %%@
a low surface brightness component underlying the starburst that could be attributed to a disk-like structure inside %%@
the galaxy. We have performed an exponential law fitting to the profiles, following the expression:
\begin{eqnarray}
I=I_o \exp(- r / \alpha),
\end{eqnarray}
which describes a typical disk structure: $I_o$ is the central intensity and
$\alpha$ is the scale length. The fits were performed between $r_{min}$ = 14$\arcsec$ and $r_{max}$ = 22$\arcsec$ and %%@
are plotted over each profile with a dashed line. The upper limit was chosen in order to avoid contamination by the %%@
bright star at the SW of the galaxy (see Figure~\ref{figR}); its effect on the surface brightness can be observed as %%@
the peak at a radius of 24$\arcsec$. Thus, it introduces an additional uncertainty to the linear fits to the %%@
underlying stellar emission. The fitting structural parameters are indicated in Table~\ref{perfiles}. 
Remark that an exponential intensity profile is typical not only for disks but also for low-mass spheroids, so it %%@
should not be taken as evidence that the 3D geometry of IRAS 08339+6517 is approximated best by a disk. Note that the %%@
fit for the $U$ filter has highest error because the image is not deep enough to reach the faint surface brightness %%@
level needed to perform a good analysis. The fits can be used to determine the relative contribution of the %%@
underlying component and the burst (we will designate as \emph{burst} the luminosity that can not be explained by the %%@
underlying component) to the luminosity at different radii. The relative contribution of the underlying component at %%@
several radii is also indicated in Table~\ref{perfiles}: it is only ~2\% at the central areas but equal to the %%@
contribution of the burst at $\sim$10$\arcsec$. From $11\arcsec$ it dominates the total luminosity of the galaxy. The %%@
burst to total luminosity ratio is $\sim$85\%.

\begin{figure}[t!]
\centering
\includegraphics[angle=0,width=0.9\linewidth]{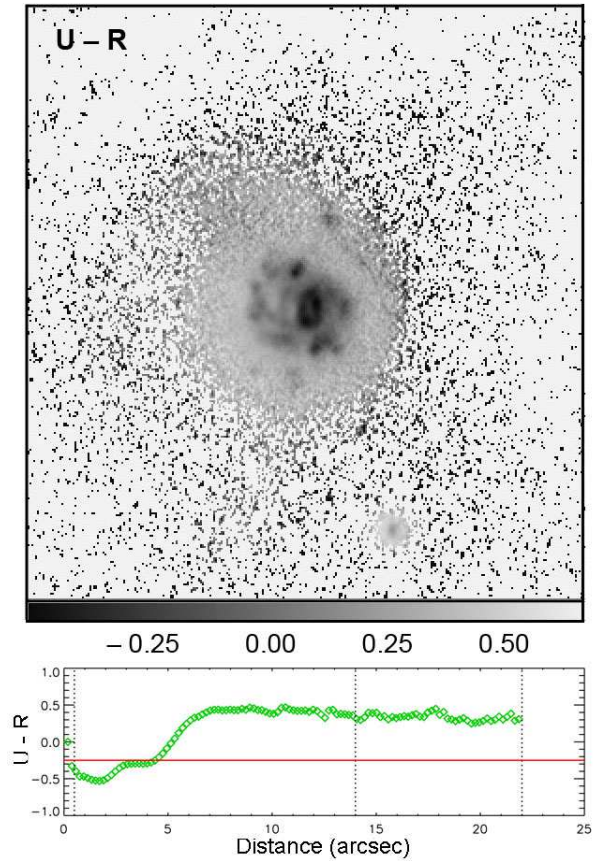}
\caption{\small{$U-R$ color map of IRAS 08339+6517, revealing a probable inhomogeneous distribution of the dust in %%@
the inner areas of the galaxy. Remark the blue ring-like structure around the nucleus of the galaxy. We also show the %%@
$U-R$ color profile for IRAS 08339+6517. Note the color excess for radii $\leq 1\arcsec$. The red horizontal line %%@
indicate the average color, $U-R = -0.25$.}}
\label{mapur}
\end{figure}

The variations of the radial color profiles also indicate the different stellar populations present in the main %%@
galaxy. Its central areas show bluer colors, but for radii $\geq 14\arcsec$ the color profiles have no gradients and %%@
show va/-lues redder to the ones derived from the integrated photometry. Surprising, we note a central color excess %%@
in the $U-B$ profile inside radius $\sim 2\arcsec$. It does not seem to be artificial because its size is larger than %%@
the seeing ($\sim 0.8\arcsec$). In Figure~\ref{mapur} we present an $U-R$ color map of IRAS 08339+6517, as well as %%@
the $U-R$ color profile. The color map shows a ring-structure with bluer colors, that includes the two central bright %%@
bursts (see Figure~\ref{figHa}), around the nucleus of the galaxy. The $U-R$ color map also represents the dust %%@
distribution throughout the galaxy, that is somewhat inhomogeneous, as we previously note in \S 3.4.3. The nucleus of %%@
the galaxy seems to be dustier than its surroundings, being probably the explanation of the central color excess %%@
observed in both $U-B$ (Figure~\ref{p1}) and $U-R$ (Figure~\ref{mapur}) color profiles. This feature is not observed %%@
in the others color profiles because extinction in $U$ filter is much more effective.

We have estimated the colors of the underlying component and the burst using the fits to the light profiles derived %%@
above. They are plotted in Figure~\ref{models}. As we expected, the underlying component has significantly redder %%@
colors than the burst, supporting that it is composed by an older population. Their $U-B \sim -0.08$ and $B-V$ = 0.16 %%@
colors imply a minimal age between 200 and 300 Myr, although it rises to 700 -- 900 Myr considering the $V-R$ = 0.30 %%@
color. For the burst component we find $U-B$ = $-0.51$, $B-V$ = 0.02 and $V-R$ = 0.23, suggesting an age between 15 %%@
and 25 Myr using both STARBURST99 and PEGASE.2 models. These values are not as young as those obtained from $W$(\Ha); %%@
it may indicate that, besides the disk-like underlying structure, which has been removed, an additional contribution %%@
of older stars also exists in the central areas of the galaxy.

On the other hand, the fits in the companion galaxy were performed between $r_{min}$ = 2$\arcsec$ and $r_{max}$ = %%@
7$\arcsec$. They do not show two structural components, although deviations from the fits can be observed in the %%@
inner area and at $r\sim 3.5\arcsec$ (see Figure~\ref{p2}) where the starburst is located and accounts for $\sim %%@
10$\% of the surface brightness. It seems to be somewhat more evident in the profile derived for the $U$-filter. In %%@
general, the contribution of the starburst is diluted under the old stellar component, that accounts of 99\% of the %%@
total luminosity of this dwarf galaxy. High spatial resolution photometry in NIR \citep{N03} should be performed for %%@
a proper separation of these two stellar components. The radial color profiles are rather constants, with values %%@
similar to those derived from aperture photometry, also supporting the idea that the galaxy is mainly dominated by %%@
this old stellar population. Remember that the $U$ filter is not deep enough to reach a faint surface brightness %%@
level; its uncertainty is high for radii greatest than $r\sim 6\arcsec$.

\subsubsection{Ages from hydrogen stellar absorption lines}

\begin{figure}[t!]
\includegraphics[angle=90,width=1\linewidth]{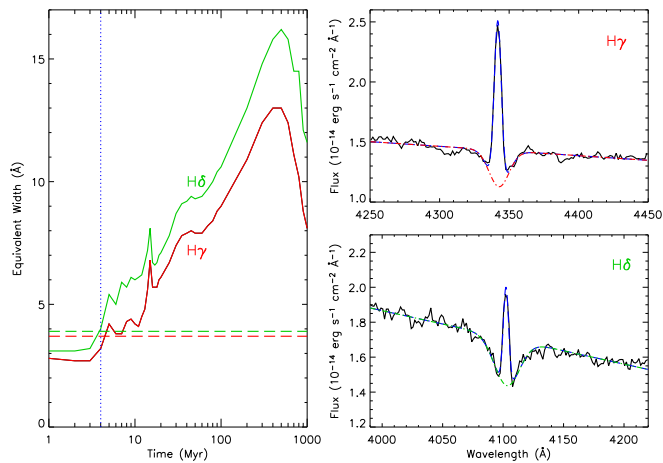}
\caption{\small{(Right) H$\gamma$ and H$\delta$ emission lines of our 12$\arcsec$ spectrum (black line) and a fit %%@
(long dashed line) using a narrow component ontop a broader absorption component (dash-dotted line). (Left) Variation %%@
of the equivalent widths of the H$\gamma$ and H$\delta$ absorption lines with the age of the burst, following the %%@
evolutionary stellar population synthesis models presented by \citet{GD99}. The dashed horizontal lines represent the %%@
equivalent widths derived from our spectrum. Note the decrement of the equivalent width for $t>$500 Myr in the %%@
models.}}
\label{hdhg}
\end{figure}

The age of the stellar populations in a starburst galaxy can be estimated analyzing the equivalent widths of hydrogen %%@
and helium absorption lines. \citet{GDL99} and \citet{GD99} presented a synthetic grid of stellar \ion{H}{i} Balmer %%@
and \ion{He}{i} absorption lines in this kind of galaxies and deve\-loped evolutionary stellar population synthesis %%@
models for instantaneous bursts with ages between 1 Myr and 1 Gyr, assuming a Salpeter IMF between 1 and 80 \Mo\ and %%@
solar metallicity. The models indicate an increment in the equivalent widths with age. This analysis has been useful %%@
in similar studies of young starbursts or blue compact dwarf galaxies  \citep[e.g.][]{GIP01}. 

We have used DIPSO software \citep{HM90} to perform a Gaussian fitting to the H$\gamma$ and H$\delta$ profiles of our %%@
12$\arcsec$ spectrum using a narrow Gaussian emission line profile ontop a broader absorption component to estimate %%@
their equivalent absorption widths. The fits are shown in Figure~\ref{hdhg}. We have obtained $W_{abs}$(H$\gamma$)= %%@
3.7 $\pm$ 0.1 \AA\ and $W_{abs}$(H$\delta$)= 3.9 $\pm$ 0.1 \AA. Comparing these values with the \citet{GD99} models, %%@
we estimate an age between 4 and 7 Myr for the stellar popu\-lation. This age is surprising similar to the one %%@
derived from the \Ha\ emission and not to that found from the analysis of the surface brightness profiles ($>$ 100 %%@
Myr). If the case of continuous star formation is considered instead of an instantaneous burst scenario, the values %%@
derived for $W_{abs}$(H$\gamma$) and $W_{abs}$(H$\delta$) will imply a larger age, $\sim$15 Myr, but still younger %%@
that 100 -- 200 Myr. \citet{GD99} models also predict than the equivalent widths of Balmer lines in absorption %%@
decrease for $\sim$500 Myr, so the derived values might also imply ages $>$1 Gyr. It would be in agreement with the %%@
estimation given by NIR colors. The analysis of H$\gamma$ and H$\delta$ equivalent widths is not useful, in this %%@
case, to constraint the age of the underlying stellar population because the result merely is a luminosity-weighted %%@
age for the mixture of stars in the central part of the galaxy. At the center of IRAS 08339+6517, the light is %%@
dominated by the young stars recently formed in the starburst, as we see both from the \Ha\ image and the fact that %%@
it accounts for $\sim$85\% of the total luminosity at this radius (the contribution of the underlying component is %%@
$\sim$15\% at $r=6\arcsec$, see Table~\ref{perfiles}). In this sense, an appropriate study of the underlying stellar %%@
population using absorption lines would require to measure them at radii r$\geq$ 15$\arcsec$, but our spectra are not %%@
able to reach these values.

%Although it is weaker, we have estimated the equivalent width of the \Hb\ absorption, obtaining $W_{abs}$(H$\beta$)= %%@
%5.0 $\pm$ 0.5 \AA. Using the \citet{GD99} models this value implies an age of 12 Myr for the underlying stellar %%@
%population, {\bf also too young}. With these data we can check if the reddening coefficient, C(H$\beta$), that we %%@
%derived in \S3.1 is correct. We calculate values of 0.23 and 0.15 using H$\gamma$/H$\beta$ and H$\delta$/H$\beta$ %%@
%ratios, respectively, that are in rather good agreement with the one we derived using the Balmer decrement, %%@
%C(H$\beta$) = 0.22 $\pm$ 0.02.   

\subsubsection{Ages from the spectral energy distribution}

\begin{figure}[t!]
\includegraphics[angle=90,width=1\linewidth]{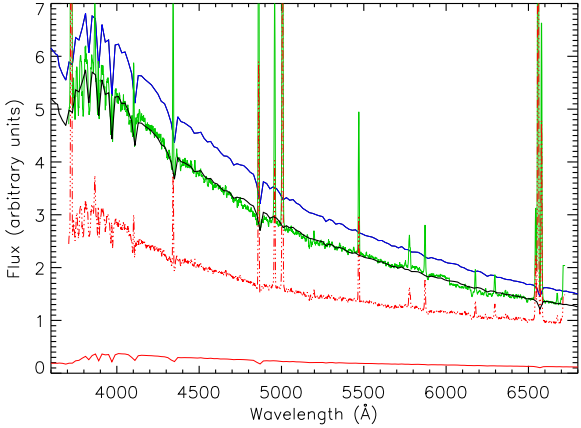}
\caption{\small{Spectrum of IRAS 08339+6517 compared with synthetic continuum spectral energy distributions obtained %%@
using the PEGASE.2 \citep{PEGASE97} code. The dotted line represents the observed spectrum uncorrected for %%@
extinction, whereas the gray/green continuous line is the extinction-corrected spectrum assuming C(\Hb)=0.22. The %%@
upper conti\-nuous line corresponds to a model with an age of 6 Myr (young population model), whereas the lower %%@
continuous one is a 140 Myr model (old population model). The shape of our observed derredened spectrum fits with a %%@
model with a contribution of 85\% for the young population and 15\% for the old population is considered (continuous %%@
black line over the galaxy spectrum). }}
\label{sed1}
\end{figure}

Finally, we have employed spectral energy distributions (SED) to constraint the ages of the stellar populations. %%@
Although this method is dependent of the interstellar extinction, our estimation of the reddening contribution using %%@
the Balmer decrement (see \S3.1) lets us to perform it avoiding the degeneracy problem between reddening and the ages %%@
of stellar populations. We have made use of the PEGASE.2 code \citep{PEGASE97} to produce a grid of theo\-retical %%@
SEDs for an instantaneous burst of star formation and ages between 0 and 10 Gyr, assuming a $Z_{\odot}$ metalli\-city %%@
and a Salpeter IMF with lower and upper mass limits of 0.1 \Mo\ and 120 \Mo. Although the grid include the ionized %%@
gas emission, we have neglected it because its contribution to the continuum is rather weak. In Figure~\ref{sed1} we %%@
show our 12$\arcsec$ extinction-corrected spectrum and the synthetic continuum spectral energy distributions derived %%@
assuming young (6 Myr) and old (140 Myr) ages. None of the individual synthetic spectra fitted our observed SED, so %%@
we constructed a model than combines 85\% of the 6 Myr model and 15\% of the 140 Myr model. We chosen these values %%@
following the relative contribution of the burst and the underlying component that we derived in the analysis of the %%@
surface brightness profiles at 6$\arcsec$ (see Table~\ref{perfiles}). We note that this combined model is in %%@
excellent agreement with the shape of our derreddened spectrum. However, we want to remark the degeneration existing %%@
in this kind of studies: several combinations of different relative contributions for old and young population models %%@
might also explain the observed spectrum. For example, a model which combines a contribution of 50\% for both 6 Myr %%@
and 100 Myr models also agrees with it. 

\begin{figure}[t]
\includegraphics[angle=90,width=1\linewidth]{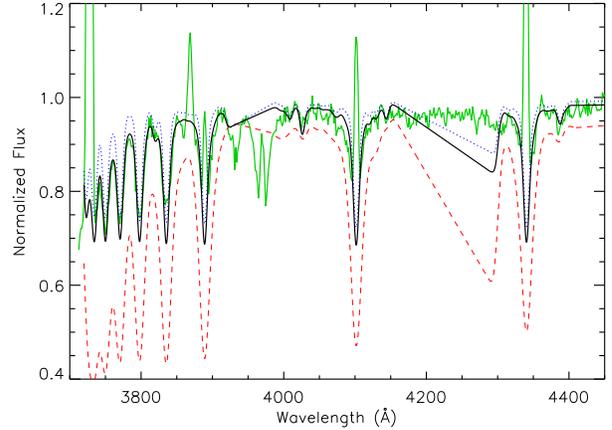}
\caption{\small{Normalized dereddened spectrum of IRAS 08339+6517 (gray/green continuous line) compared with %%@
\citet{GD99} models with 4 Myr (dotted line) and 200 Myr (dashed line) at $Z_{\odot}$ metallicity. The best fit %%@
corresponds to a model with a contribution of 85\% for the young population and 15\% for the old population %%@
(continuous black line). Note that the models only give values for the position of the absorption lines and not for %%@
all wavelengths, being the reason of the straight lines presented between 3930 and 4000 \AA\ and 4150 and 4300 \AA.
The straight lines are artifacts connecting gaps in the dataset of the models.}}
\label{sed2}
\end{figure}

The agreement between observed and synthetic spectrum is also found using the \ion{H}{i} Balmer and \ion{He}{i} %%@
absorption lines derived from \citet{GD99} models. In Figure~\ref{sed2} we plot our derredened normalized spectrum in %%@
the 3700 -- 4450 \AA\ range compared with their $Z_{\odot}$ models with ages of 4 and 200 Myr. The spectral %%@
resolution of the models were degraded to that of our observed spectrum. Again, a good fit is found when a combined %%@
model with a contribution of 15\% for the old population and 85\% for the young population is considered.

In conclusion, the young stellar population in IRAS 08339+6516 (the most recent starburst) has an age between 4 and 6 %%@
Myr, and it is located in the inner areas of the galaxy (radii $<$ 6$\arcsec$). It is superposed to a more evolved %%@
stellar population with an age not younger than 100 -- 200 Myr, that fits a disk-like profile, and probably formed in %%@
early bursts. The NIR colors suggest that an older stellar population, with ages larger than 1 -- 2 Gyr, also exists %%@
in IRAS 08339+6517. Its companion galaxy is practically dominated by an old population with age $>$ 250 Myr, although %%@
a recent starburst of around 6 Myr is found in its external areas.

\subsection{The star formation rate}

\begin{table*}[t!]\centering
 \caption{Star formation rates (SFR) derived for IRAS 08339+6517 using different relations (see \S4.2). All the %%@
luminosities were calculated assuming a distance of 80 Mpc for the galaxy.}
 \label{multi}
 \footnotesize
 \begin{tabular}{lccccc}
 \noalign{\smallskip}
   \hline\hline
   \noalign{\smallskip}
Range &  Luminosity    & Units   & Value     &   SFR (M$_{\odot}$ yr$^{-1}$)   &  Calibration \\
    \hline
    \noalign{\smallskip}
%X-ray & F$_{0.3-3.5\ keV}$       & (1.69 $\pm$ 0.32)$\times$ 10$^{-13}$ & NO & 1990 ApJ 72 567 \\
%      & F$_{0.2-2.0\ keV}$ ROSAT & (1.62 $\pm$ 0.25)$\times$ 10$^{-13}$ & NO & 2000WGA....\\ 
%X-ray & L$_{X(0.2-2.0\ keV)}$ ROSAT & erg s$^{-1}$  & (1.24 $\pm$ 0.18)$\times$ 10$^{41}$  & 27.3&2000WGA+Ranalli\\
X-ray & L$_{X(0.2-2.0\ keV)}$  & erg s$^{-1}$  & 2.81 $\times$ 10$^{41}$  & 61.8 & \citet{RCS03}\\
%	  & L$_X$ Einstein IPC       & 1.0 $\times$ 10$^{41}$ erg s$^{-1}$  &  NO & Margon et al. 1998\\
%	  & L$_X$ ROSAT              & 2.8 $\times$ 10$^{41}$ erg s$^{-1}$  &  NO & Steven et al. 1998\\

%\noalign{\smallskip}
%UV   &  F$_{Ly\alpha}$  & erg s$^{-1}$ cm$^{-2}$  & 4 $\times$ 10$^{-11}$   &  ?  & Gonzalez Delgado p714 \\
%     &  F$_{Ly\alpha}$  & erg s$^{-1}$ cm$^{-2}$  & 5.6 $\times$ 10$^{-14}$   &  ?  & \citet{KMH98} p16\\

\noalign{\smallskip}
 Optical  &  L$_{H\alpha}$   & erg s$^{-1}$ & (12.0 $\pm$ 0.6) $\times\ 10^{41}$ & 9.5 $\pm$ 0.5 & \citet{K98}\\
%         &  F$_{H\alpha}$   &  (1.60 $\pm$ 0.07)$\times\ 10^{-12}$  &   NO & This work  \\
%         &  L$_U$           & L$_{\odot}$  & (1.16 $\pm$     ) $\times$ 10$^{11}$ & 1.0       & Cram et al. 1998\\
          &  L$_B$           & L$_{\odot}$  & (6.61 $\pm$ 0.24) $\times$ 10$^{10}$ & 1.92 $\pm$ 0.07 & \citet{G84} \\
		  &  L$_{B, UC}$     & L$_{\odot}$  & (8.95 $\pm$ 0.30) $\times$ 10$^9$    & 1.1 $\pm$ 0.1 & \citet{Cal01} \\
          &  L$_{[\ion{O}{ii}]}$& erg s$^{-1}$ &  1.21 $\times$ 10$^{42}$  &  13       & \citet{K98}  \\
          &                  &              &                           &  8.9      & \citet{KGJ04} \\
  
\noalign{\smallskip}
%  &   F$_{FIR}$     &   2.815  $\times$ 10$^{-10}$        &   NO           & Datos IRAS + TW  \\
Far-Infrared  &   L$_{FIR}$     & L$_{\odot}$ &   5.63 $\times$ $10^{10}$  &  9.7       & \citet{K98}   \\
			  &   L$_{12\mu m}$ & L$_{\odot}$ &   1.78 $\times$ $10^{10}$  & 11.6       & \citet{RSVB01} \\
  			  &   L$_{60\mu m}$ & L$_{\odot}$ &   3.90 $\times$ 10$^{10}$  &  9.2       & \citet{Condon92}  \\
%			  &   L$_{60\mu m}$ & W Hz$^{-1}$ &   4.60 $\times$ 10$^{24}$  &  9.2       & C92  \\
			  
\noalign{\smallskip}
%Radio       &   F$_{1.42 GHz}$ (\ion{H}{i} 21 cm) &  3.68 $\pm$ 0.46 Jy km s$^{-1}$ TODO  & NO & \citet{CSK04} \\
Radio         &   L$_{1.4\ GHz}$  & W Hz$^{-1}$ &  2.57 $\times$ 10$^{22}$&  6.4    & \citet{CCB02}   \\
%			  &   L$_{1.49\ GHz}$ & W Hz$^{-1}$ &  2.45 $\times$ 10$^{22}$& \nodata &  Condon et al 1990 \\
%\noalign{\smallskip}
%    *** FLUJOS   &   F$_{1.4\ GHz}$  &  mJy        &  33.56                   & NO    &  Condon 90 (extracted)  \\
%			  &   F$_{1.4\ GHz\ thermal}$ & mJy &   1.94                   & NO         &  Dopita 2002 \\
\noalign{\smallskip}
 \hline
 \hline
 \end{tabular}
\end{table*}

We have derived the star formation rate (SFR) of IRAS 08229+6517 using empirical calibrations in different wavebands. %%@
The SFR is a key parameter for characterizing the formation and evolution of the galaxies. However, even in the %%@
well-observed galaxies of the local Universe, it remains quite uncertain because different methods yield different %%@
values. Much of this uncertainty is related to the unknown contribution of the dust obscuration in and around the %%@
star-forming regions.

One of the most extended methods of deriving the SFR is the use of hydrogen recombination line fluxes, especially the %%@
\Ha\ flux. Since the flux in a hydrogen recombination line is proportional to the number of ionizing photons produced %%@
by the hot stars (which is also proportional to their birthrate), the SFR can be easily derived. The most recent %%@
calibration for starbursts is that derived by \citet{K98}.
%\begin{eqnarray}
%SFR_{H\alpha} ({\rm M_{\odot}\ yr^{-1}}) = 7.94 \times 10^{-42}L_{H\alpha} (\rm erg\ \rm{s^{-1}}),
%\end{eqnarray}
We have used this calibration to estimate the SFR in IRAS 08339+6517 from our \Ha\ image, as we previously indicated %%@
in \S 3.3, obtaining SFR$_{H\alpha}$ = 9.5 \Mo\ yr$^{-1}$ (see Table~\ref{multi}). Note that we have corrected the %%@
\Ha\ flux for both reddening and [\ion{N}{ii}] contamination. 

Many of the problems found in deriving the SFR from optical data can be avoided by measuring the far-infrared (FIR) %%@
and sub-millimeter spectral energy distribution. These are determined by the reradiation by the dust of energy %%@
absorbed in the visible and UV regions of the spectrum. Assuming that the dust completely surrounds the star forming %%@
regions, it acts as a bolometer reprocessing the luminosity produced by the stars.  
%Using theoretical stellar flux %distributions and evolutionary models, 
%as STARBURST99 \citep{L99}, the SFR can also be derived. 
%\citet{K98} gives the following correlation between the SFR and the far-infrared flux:
%\begin{eqnarray}
%SFR_{FIR}= 4.5 \times 10^{-44} L_{FIR},
%\end{eqnarray}
%where $L_{FIR}$ is obtained using
%\begin{eqnarray}
%F_{FIR} = 1.26 \times 10^{-11} \big( 2.58 f_{60} + f_{100} \big),
%\end{eqnarray}
%being $f_{60}$ and $f_{60}$ the flux densities for 60$\mu$m and 100$\mu$m (in Jy), and the conventional expression %%@
%between flux and luminosity, $L = 4\pi D^2 F$. 
We have applied the \citet{K98} correlation between the SFR and the FIR flux using the \emph{IRAS} satellite data for %%@
IRAS 08339+6517, $f_{60}$=5.90 Jy and $f_{100}$=6.50 Jy \citep{Mo90}, deriving SFR$_{FIR}$ = 9.7 \Mo\ yr$^{-1}$. This %%@
value is in excellent agreement with the estimation derived from the \Ha\ luminosity, indicating that the value for %%@
the extinction we have adopted seems to be fairly appropriate. We have also used others correlations between SFR and %%@
60$\mu$m \citep{Condon92} and 15$\mu$m \citep{RSVB01} luminosities (the last one assuming $L_{15\, \mu m}\ \sim\ %%@
L_{12\, \mu m}$), finding similar values (see Table~\ref{multi}).

%Roussel et al. 2001, A\&A 372, 427 give a relation between the 15 $\mu$m luminosity and SFR:
%\begin{eqnarray}
%SFR_{15\, \mu m}  \sim 6.5 \times 10^{-9} L_{15\, \mu m} (L_{\odot}), 
%\end{eqnarray}
%assuming $L_{15\, \mu m}\ \sim\ L_{12\, \mu m}$ = 1.78 $\times$ 10$^{10}$ $L_{\odot}$, we find $SFR_{15\, \mu m}$ %%@
%$\sim$ 11.6 \Mo\ yr$^{-1}$. Condon (1992) derives the following relation between SFR and L$_{60\, \mu m}$:
%\begin{eqnarray}
%SFR_{60\, \mu m}  \sim 1.96 \times 10^{-24} L_{60\, \mu m} \ ({\rm W\ Hz^{-1}}), 
%\end{eqnarray}
%obtaining for our galaxy SFR$_{60\, \mu m}$ = 9.2 \Mo\  yr$^{-1}$.

The radio continuum flux can be also used as a star formation indicator. Nearly all of the radio luminosity from %%@
galaxies without a significant active galaxy nuclei (AGN) can be traced to recently formed massive ($M\geq$ 8 \Mo) %%@
stars \citep{C92}. Ten percent of the continuum emission at 1.4 GHz is due to free-free emission from extremely %%@
massive main-sequence stars (thermal emission) and almost 90\% is synchrotron radiation from relativistic electrons %%@
accelerated in the remmants of core-collapse supernovae (non-thermal emission). As the stars that contribute %%@
significantly to the radio emission have lifetimes $\tau\leq\ 3\ \times\ 10^7$ yr and the relativistic electrons have %%@
lifetimes $\tau\leq\ 10^8$ yr, the current radio luminosity is nearly proportional to the rate of massive star %%@
formation during the past $\tau\leq\ 10^8$ yr \citep{CCB02}. 
%\begin{eqnarray}
%%%%SFR_{1.4\,\rm{GHz}} (M\ > 5M_{\odot}) \sim 2.17 \times 10^{-22} L_{1.4\,\rm{GHz}}
%%%%SFR_{1.4\,\rm{GHz}} (M\ > 0.1M_{\odot}) \sim 1.19 \times 10^{-21} L_{1.4\,\rm{GHz}}
%SFR_{1.4\,\rm{GHz}} \sim 2.5 \times 10^{-22} L_{1.4\,\rm{GHz}} \ ({\rm W\ Hz^{-1}}).
%\end{eqnarray}

We have estimated the SFR from the 1.4 GHz lumino\-sity for IRAS 08339+6517 using the \citet{CCB02} correlation. The  %%@
1.4 GHz luminosity was derived from the 1.49 GHz luminosity \citep{Condon90} applying the relation between $L_{1.49\, %%@
\rm{GHz}}$ and $L_{1.4\,\rm{GHz}}$ given by \citet{Condon91}. The obtained value, SFR$_{1.4\,\rm{GHz}}$ = 6.4 \Mo\ %%@
yr$^{-1}$, is a little lower than those derived from previous estimates but, as it was commented before, this rate %%@
corresponds to a larger time period than the \Ha\ or FIR SFR indicators. Using the reddening-corrected \Ha\ flux and %%@
the expression given by \citet{DPKC02}, we can also derive the thermal flux at 1.4 GHz for the galaxy, %%@
$F_{1.4\,\rm{GHz\, thermal}}$ = 1.94 mJ. It corresponds to around 6\% of the total 1.4 GHz flux, in agreement with %%@
the average value found in starburst galaxies \citep{Condon92}.

%Condon et al. (1990) ApJSS 73, 359 presented VLA 1.49 GHz map of the sources of the IRAS Bright Galaxy Samples. IRAS %%@
%08339+6517 was included in this study, finding L$_{1.49\ GHz}$ $\sim$ 2.45 $\times$ 10$^{22}$ W Hz$^{-1}$. Using the %%@
%relation between L$_{1.49\, \rm{GHz}}$ and L$_{1.4\,\rm{GHz}}$ shown by Condon et al. 1991, ApJ 376, 95: 
%\begin{eqnarray}
%\log \, L_{1.4\,\rm{GHz}} \sim\ \log \, L_{1.49\,\rm{GHz}}\ + \ 0.7 \log \big( 1.49 / 1.40 \big),
%\end{eqnarray}
%we derived L$_{1.4\ GHz}$ $\sim$ 2.57 $\times$ 10$^{22}$ W Hz$^{-1}$ and SFR$_{1.4\,\rm{GHz}}$ = 6.4 \Mo\ yr$^{-1}$. %%@
%Using the reddening-corrected \Ha\ flux and the expression given by Dopita et al 2002 ApJS 143:
%\begin{eqnarray}
%F_{1.4\,\rm{GHz\, thermal}} ({\rm mJy}) = 1.21 \times 10^{12} F_{H\alpha},
%\end{eqnarray}
%we can derive the thermal flux at 1.4 GHz for the galaxy, F$_{1.4\,\rm{GHz\, thermal}}$ = 1.94 mJ, that is, around %%@
%0.06 \% of the total 1.4 GHz flux, in agreement with the average value found in starburst galaxies (Condon 1992**)

Assuming that $L_B$ = 6.61 $\times$ 10$^{10}$ $L_{\odot}$ and using the relation given by \citet{G84} we derive a %%@
SFR$_B = 1.9 M_{\odot}$ yr$^{-1}$. \citet{Cal01} give a correlation between the SFR of the starburst and the $B$ %%@
luminosity of the host galaxy. Using the structural parameters derived for IRAS 08339+6517 (see %%@
Table~\ref{perfiles}), we find that the total $B$-luminosity of the underlying component is $\sim 15\%$ of the total, %%@
$L_{B,\,UC} = 8.95\times 10^{9}\ L_{\odot}$. Applying the \citet{Cal01} correlation, we derive SFR$_B$ = 1.1 \Mo\ %%@
$yr^{-1}$. The SFR derived from the blue luminosity corresponds to the past few billion years, whereas those derived %%@
from the \Ha\ or FIR fluxes indicate the current ($<10^7$ yr) SFR. 

We can also use the relation given by \citet{K98} between the SFR and the luminosity of [\ion{O}{ii}].
%\begin{eqnarray}
%SFR_{[\ion{O}{ii}]} = 1.40 \times 10^{-41} L_{[\ion{O}{ii}]}.
%\end{eqnarray}
Using the \Ha\ luminosity derived from our images for the entire galaxy (see Table~\ref{table2}) and the %%@
[\ion{O}{ii}]/\Ha\ ratio from our spectra ([\ion{O}{ii}]/\Ha\ = 1.01 for the 12$\arcsec$ aperture), we find $SFR_{\rm %%@
[\ion{O}{ii}]}$ = 13 \Mo\ yr$^{-1}$, considering the low-limit of the \citet{K98} calibration, corresponding to blue %%@
emission-line galaxies. A more realistic relation between the SFR and the luminosity of [\ion{O}{ii}], that take into %%@
account the oxygen abundance and ionization parameter of the ionized gas, was presented by \citet{KGJ04} (their eq. %%@
15). Applying this relation, we obtain $SFR_{\rm [\ion{O}{ii}]}$ = 8.9 \Mo\ yr$^{-1}$.

Global soft X-ray luminosity may seem to serve as an indicator for the SFR in star-forming galaxies.
In the last years, several authors have tried to find a correlation between the soft X-ray emission and the SFR. %%@
%Ranalli et al. (2003) obtained the following relation:  
%\begin{eqnarray}
%SFR_{L_X} = 2.2 \times 10^{-40} L_{X\ (0.5-2.0\ keV)}.
%\end{eqnarray}
Using the \citet{RCS03} expression and the $L_X$ value from the \emph{ROSAT} Satellite \citep{SS98}, L$_{0.2-2.0\ %%@
keV}=2.81\times 10^{41}$ erg s$^{-1}$, we derived $SFR_{L_X}=$ 61.8 \Mo\ $yr^{-1}$, a very high value compared with %%@
that found using the others calibrations. \citet{SS98} find that the X-ray luminosities of WR galaxies are %%@
considerably higher than those of the other galaxies with the same $B$ luminosities, consequence of the higher %%@
occurrence of superbubbles in WR galaxies. Superbubbles are hollow cavities with size of the order of kpc, expanding %%@
with velocities between 25 and 150 km s$^{-1}$ produced by the combination action of supernova explosions and stellar %%@
winds \citep{L94}. We can estimate the supernova rate, $\nu_{SN}$, using the expression (7) in \citet{Kewley00} that %%@
assumes a Salpeter IMF with lower and upper mass limits of 0.1 and 100 \Mo\ and a minimum initial mass for supernova %%@
detonation of 8 \Mo:
\begin{eqnarray}
\nu_{SN} = 7.42 \times 10^{-3}\times {\rm SFR}.
\end{eqnarray}
Assuming a SFR = 9.5 \Mo\ yr$^{-1}$, we find a supernova rate of $\nu_{SN}\ \sim$ 0.07 yr$^{-1}$, i.e., a supernova %%@
explosion each 14 years, a bit higher than the value  usually found in starburst galaxies [around 0.02 yr$^{-1}$ %%@
\citep{Kewley00}]. Thus, the relation given by \citet{RCS03} between SFR and the X-ray luminosity seems to be not %%@
suitable for very young starbursts.

The gas depletion timescale is defined as:
\begin{eqnarray}
\tau_{gas} = 1.32 \times M_{\rm H \, I}\ / \ {\rm SFR},
\end{eqnarray}
\citep{SCM03}, indicating the number of years that a galaxy may continue to form stars at its current rate. The %%@
factor 1.32 was introduced to account for He. For IRAS 08339+6517 we obtain $\tau_{gas}$ = 0.15 Gyr (0.68 Gyr %%@
considering that the mass of the \ion{H}{i} tidal tail). It is a very low value, indicating the starbursting nature %%@
of the galaxy.

In conclusion, the SFR derived using the \Ha, [\ion{O}{ii}], FIR, 15$\mu$m, 60$\mu$m and 1.4GHz luminosities are in a %%@
very good agreement and give a SFR $\sim$ 9.5 \Mo\ yr$^{-1}$. The value of the SFR found using the $B$-luminosity of %%@
the underlying population suggests that the star formation in the last 100 Myr has not been as high as the present %%@
rate. Furthermore, the high SFR derived using X-ray luminosities implies a strong massive star formation in the last %%@
few Myr, supporting a high rate of supernova explosions and the possible detection of the WR stars.

\subsection{WR population}

\begin{figure}[t!]
\includegraphics[angle=270,width=1\linewidth]{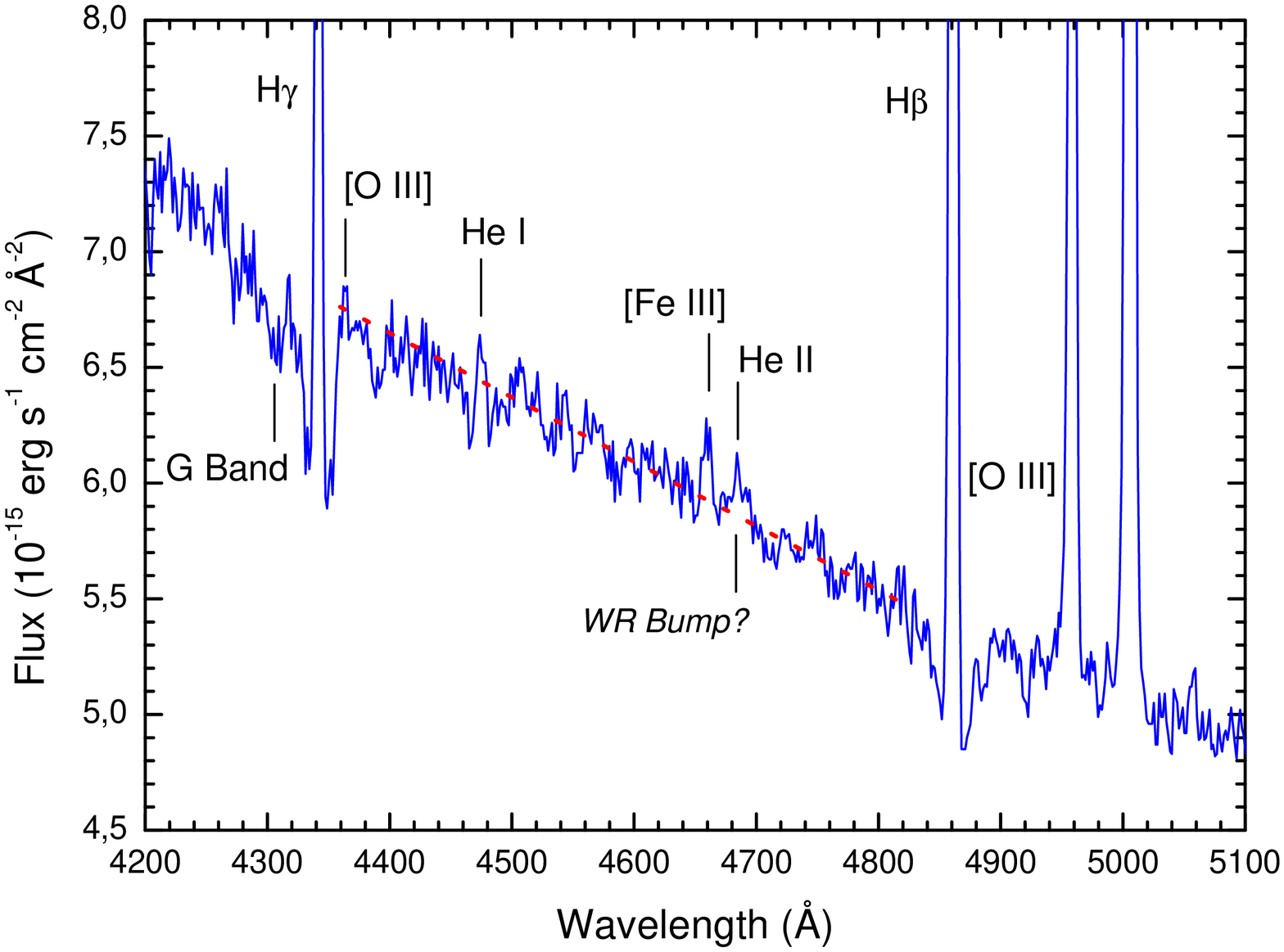}
\caption{\small{Part of the spectrum of IRAS 08339+6517 between 4300 and 5100 \AA\ showing the \ion{He}{ii} %%@
$\lambda$4686 emission line associated with Wolf--Rayet stars. The dotted line indicates a linear fit to the %%@
continuum.}}
\label{wr}
\end{figure}

IRAS 08339+6517 was included in the X-ray study of WR galaxies performed by \citet{SS98} because previous authors %%@
\citep{C91,GDL98} suggested that it has recently passed through the WR galaxy phase. However, our deep optical %%@
spectrum reveals that WR stars could still be present in the starburst. In Figure~\ref{wr} we show part of the %%@
spectrum of the central object of IRAS 08339+6517 (knot \#1) between 4200 and 5100 \AA. It shows a very weak bump %%@
between 4660 and 4700 \AA\ and an emission line at $\lambda$4685.4 that could correspond to the \ion{He}{ii} %%@
$\lambda$4686 emission line. If this assumption is correct it would indicate the first detection of WR stars in this %%@
starburst galaxy. The UV spectrum of IRAS 08339+6517 presented by \citet{MAM88} showed a very weak emission line %%@
around 1640 \AA\ that could be attributed to \ion{He}{ii} $\lambda$1640, line also associated with WR stars. However, %%@
these authors ruled this out because they do not detect the \ion{He}{ii} $\lambda$4686 in their optical spectrum.

% L(HeII 4686)=5.284E38

Although the \ion{He}{ii} $\lambda$4686 emission line is weak, we have used the evolutionary synthesis models for O %%@
and WR populations in young starburst of \citet{SV98} to perform a tentative estimation of the WR/(WR+O) ratio. %%@
Assuming that all the contribution of the \ion{He}{ii} $\lambda$4686 emission line comes from WNL stars and %%@
considering a luminosity of $L$(WNL 4686) = 1.7$\times10^{36}$ erg s$^{-1}$ for a WNL star \citep{VC92}, we find %%@
around 310 WNL stars in this burst. To derive the WR/(WR+O) ratio, the contribution of the WR stars to the total %%@
ionizing flux must be considered to obtain the total number of O stars. Assuming a luminosity of $L$(\Hb) = %%@
4.76$\times10^{36}$ erg s$^{-1}$ for a O7V star \citep{VC92} and $\eta\equiv$ O7V/O = 0.25 for an age of around 4.5 %%@
Myr \citep{SV98}, we derive around 10700 O stars. This implies a WR/(WR+O) ratio of 0.03 in knot \#1. An identical %%@
value is found using the calibration between the WR/(WR+O) ratio and the flux of the WR bump given by \citet{SV98} %%@
(their equation 17).

%We wish to stress aperture effects in the detection of WR stars. 
The weak \ion{He}{ii} $\lambda$4686 emission line is only detected in the spectra of knot \#1 (1$\arcsec$ aperture) %%@
and in the 2.5$\arcsec$ aperture but it is not found in any of the other spectra of the same galaxy. We suspect that %%@
maybe knot \#1, which hosts the youngest and most powerful burst in the system, is the only area where a substantial %%@
population may be present. 
%When a larger aperture is extracted, the weak emission line is diluted by the continuum flux. 
Thus, aperture effects and the position of the slit can play an important role in the detection of WR features, as %%@
some authors have previously pointed out \citep{HGJL99,LSER04a,LSER04b}. 

%As WR stars are only located in the youngest \ion{H}{ii} regions, the exact location of these WR-rich areas is %%@
%therefore essential to analyze this sort of massive stars in starburst galaxies.

However, although the \ion{He}{ii} $\lambda$4686 emission line seems to really exist in knot \#1, deeper %%@
spectroscopical data will be necessary to confirm that it is broad and/or that the WR bump undoubtedly exists. Only  %%@
under these facts the classification of IRAS 08339+6517 as \emph{Wolf-Rayet galaxy} can be established beyond any %%@
doubt.

\begin{figure}[t!]
\includegraphics[angle=270,width=1\linewidth]{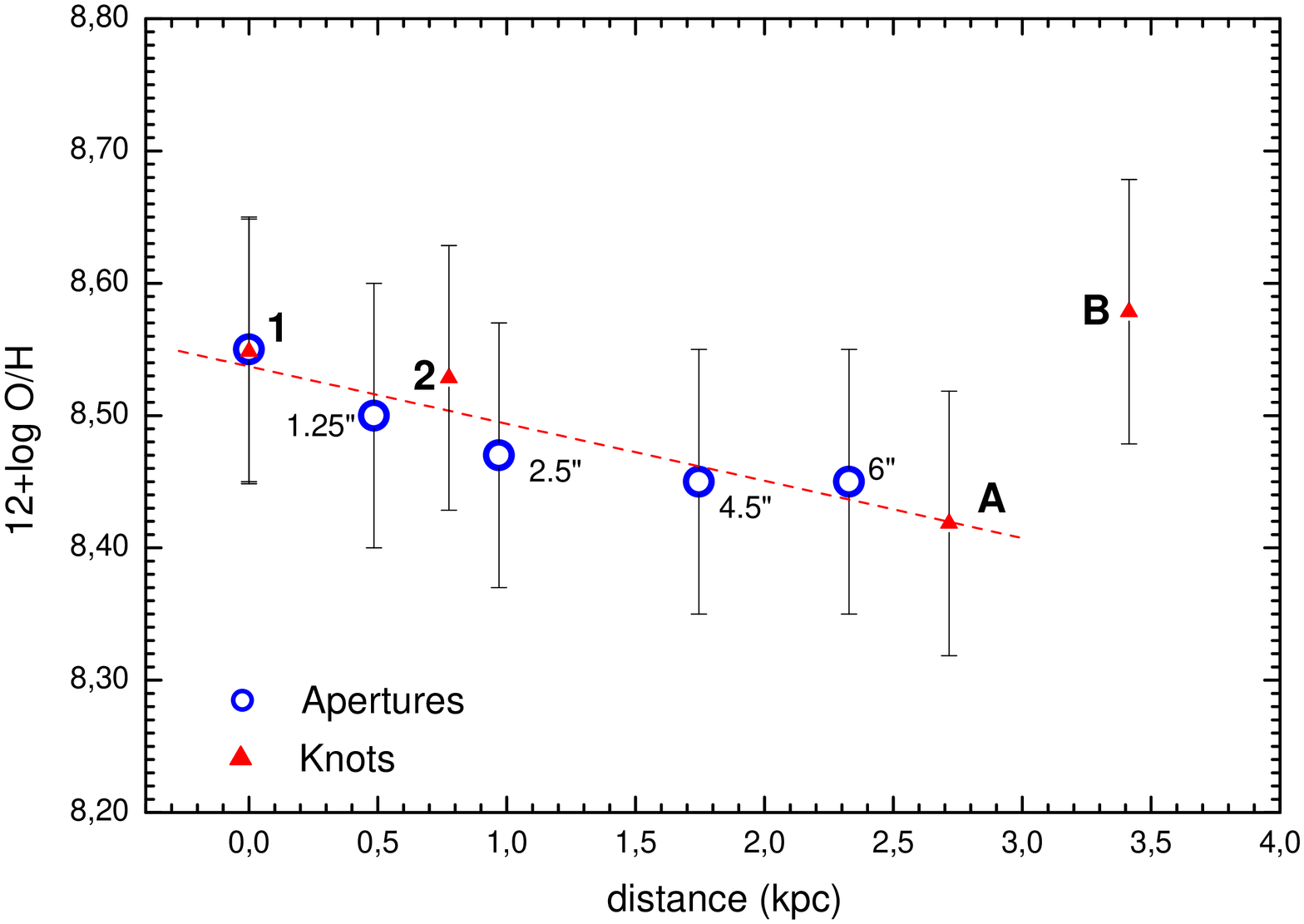}
\caption{\small{Relation between metallicity and distance from the center of IRAS 08339+6517 for the observed knots %%@
(triangles) and different aperture sizes (open circles). We have considered the oxygen abundances derived using the %%@
calibration by \citet{P01}. The dotted line is a linear fit to all the data except knot B,  which shows the higher %%@
O/H value. The li\-near fit suggests an abundance gradient along the disk.}}
\label{radial}
\end{figure}

\subsection{The nature of IRAS 08339+6517}

\subsubsection{Radial dependence of metallicity and reddening}

We have analyzed the radial dependence of the derived metallicity of the eight apertures traced in IRAS 08339+6517 %%@
(four apertures corresponding with knots and four apertures with different sizes and centered on the nucleus, see %%@
Figure~\ref{rendija}), which is shown in Figure~\ref{radial}. The oxygen abundance obtained from the empirical %%@
calibration of \citet{P01} (see Tables~\ref{table5} and \ref{radios}) was assumed for each aperture.  

Except for knot B, the oxygen abundances derived from the nucleus [knot \#1 with 12 + log(O/H) = 8.55] to the outer %%@
regions [knot A with 12 + log(O/H) = 8.42] is continuously decreasing. Thus, despite the uncertainties, a possible %%@
presence of a weak abundance gradient along the disk of IRAS 08339+6517 (around $-$0.043 dex kpc$^{-1}$) is detected. %%@
The average gradient in spiral galaxies is around $-$0.06 dex kpc$^{-1}$ \citep{ZKH94}. 

%sale regresion 12+log/O/H)=8.54 -0.043 dex/kpc

\begin{figure}[t!]
\includegraphics[angle=270,width=1\linewidth]{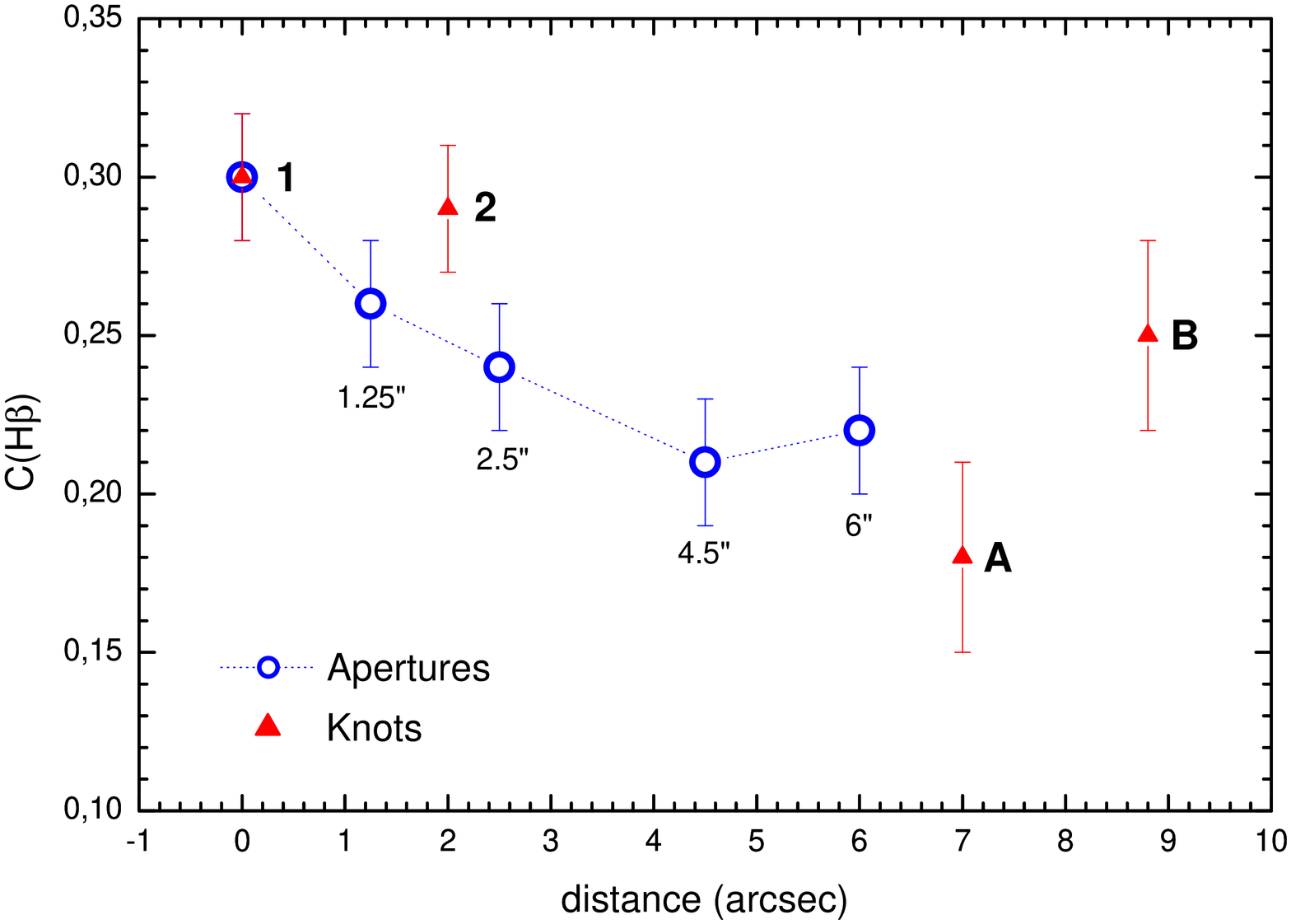}
\caption{\small{Relation between the reddening coefficient, C(\Hb), with distance from the center of IRAS 08339+6517. %%@
Observed knots are plotted with triangles and the different aperture sizes with open circles connected with a dotted %%@
line.}}
\label{radial3}
\end{figure}

Besides the oxygen metallicity, the $W_{abs}$ calculated from the Balmer decrement increases with the aperture size, %%@
confirming again the existence of an evolved underlying stellar population. The C(\Hb) derived for different %%@
apertures inside IRAS~08339+6517 also indicate an apparent slight gradient between the nucleus and the external zones %%@
of the galaxy (see Figure~\ref{radial3}). This may explain the differences found by \citet{GDL98} in the color excess %%@
between two different aperture sizes. For the nucleus, they estimated $E(B-V)$=0.19 (from the spectrum obtained with %%@
a $1.7\arcsec\times 1.7\arcsec$ aperture using the Goddard High-Resolution Spectrograph on the \emph{Hubble Space %%@
Telescope}), similar to the value we derive for our $1\arcsec\times 1\arcsec$ aperture, $E(B-V)$=0.20$\pm$0.01 %%@
[C(\Hb)=0.30$\pm$0.02]. For the external zones, they derived $E(B-V)$=0.10 (using the $12\arcsec$ circular aperture %%@
with the \emph{Hopkins Ultraviolet Telescope}), which corresponds with the one we obtained for our $12\arcsec\times %%@
1\arcsec$ aperture, $E(B-V)$=0.12$\pm$0.01 [C(\Hb)=0.22$\pm$0.02]. The effect of the aperture size on derived %%@
integrated properties of galaxies has been recently studied by \citet{KJG05}, who concluded that the difference %%@
between the nuclear and global derived values for metallicity, extinction and SFR is substantial: for example, for a %%@
late-type spiral galaxy, the metallicity of the nucleus is $\sim$0.14 dex greater than the global metallicity of the %%@
galaxy.

\begin{table}[t!]\centering
 \caption{Principal properties derived for the ionized gas in IRAS 08339+6517 using different aperture sizes.}
 \label{radios}
 \footnotesize
 \begin{tabular}{cccc}
 \noalign{\smallskip}
   \hline\hline
   \noalign{\smallskip}
Aperture ($\arcsec$) &  C(\Hb)   &   W$_{abs}$(\ion{H}{i}) (\AA) &  12+log(O/H)  \\
    \hline
    \noalign{\smallskip}
	1 $\times$ 1   & 0.30 $\pm$ 0.02  &  1.1 $\pm$ 0.1  &  8.55 $\pm$ 0.10  \\
	2.5 $\times$ 1 & 0.26 $\pm$ 0.02  &  1.4 $\pm$ 0.1  &  8.50 $\pm$ 0.10 \\
	 5 $\times$ 1  & 0.24 $\pm$ 0.02  &  1.5 $\pm$ 0.1  &  8.47 $\pm$ 0.10  \\
	  9 $\times$ 1 & 0.21 $\pm$ 0.02  &  1.6 $\pm$ 0.1  &  8.45 $\pm$ 0.10 \\
     12 $\times$ 1 & 0.22 $\pm$ 0.02  &  1.8 $\pm$ 0.1  &  8.45 $\pm$ 0.10 \\
 \hline
 \hline
 \end{tabular}
\end{table}

\subsubsection{Knot B: a TDG or the remnant of a merger?}

\begin{figure*}[t!]
\centering
\includegraphics[angle=90,width=1\linewidth]{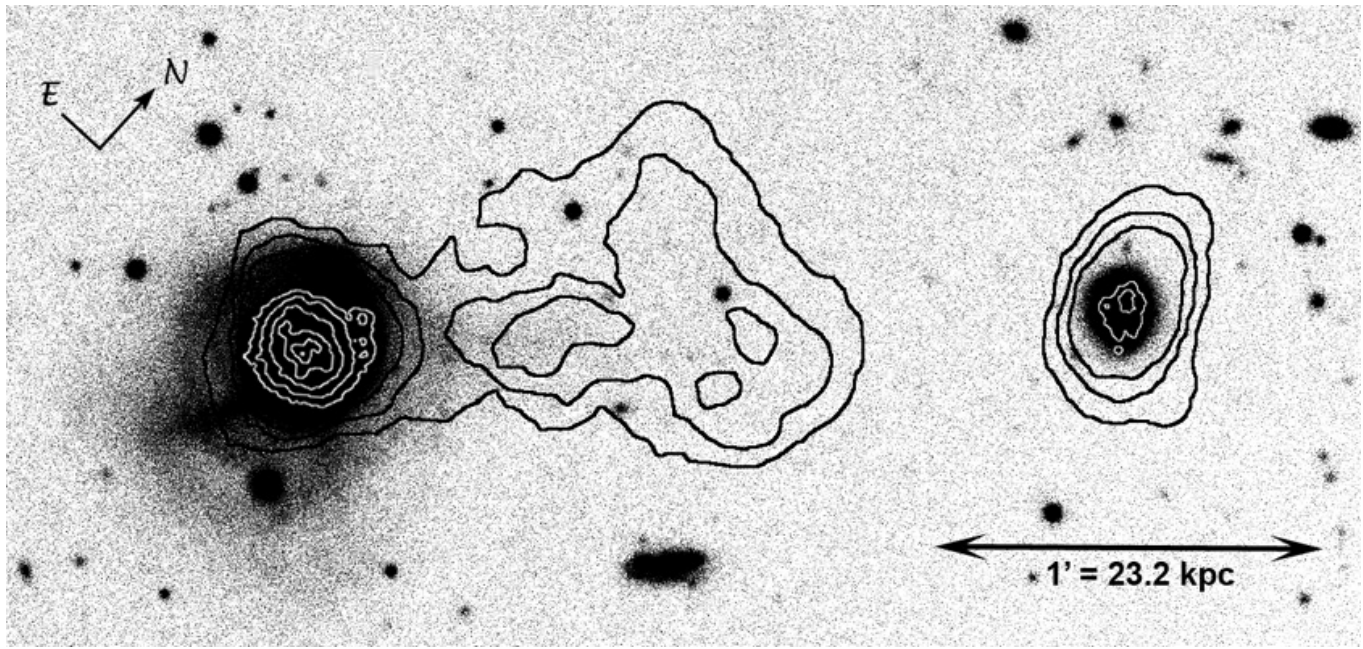}
\caption{\small{Deep $R$ image of IRAS 08339+6517 and its companion galaxy showing the faintest details. Our %%@
continuum-subtracted \Ha\ image (white contours) and the \ion{H}{i} map obtained by \citet{CSK04} (black contours) is %%@
superposed. Note that the weak optical plume in the direction of the companion galaxy coincides with the \ion{H}{i} %%@
tidal tail.}}
\label{compara}
\end{figure*}

As it can be observed in Figure~\ref{radial}, the external knot B, located 3.4 kpc from the center of IRAS %%@
08339+6517, shows a high metallicity [12 + log(O/H) = 8.58], even a bit higher than that derived for the nucleus of %%@
the galaxy. Note that knots A and B are at rather similar distances from the nucleus but are located on opposite %%@
sides of the galaxy (see Figure~\ref{figHa}). Their spectra also show approximately the same S/N. Knot A seems to be %%@
a bright \ion{H}{ii} region (or a complex of \ion{H}{ii} regions) in the outer areas of the disk. 
%whereas knot B has similar metallicity than the center of the galaxy. 
Furthermore, as we can see in \S3.4.3, knot B also shows a decoupled kinematics with respect to the motion of ionized %%@
gas associated to the tidal tail, so it might not be symply an intense star-formation region in the outskirts of the %%@
disk. We consider two possible explanations to these facts:

\begin{enumerate}  
%\item It is just an intense star-formation region in the outskirts of the spiral disk.
%\item Knot B corresponds to an outflow driven by galactic winds. It would explain its similar metallicity to that of %%@
%the nucleus of the galaxy. Outflows can accelerate gas to high velocities, but they also show broader or %%@
%doubled-peaked emission lines. These signatures are not observed in none of our spectra but are detected in the %%@
%Ly$\alpha$ emission of the inner areas of the galaxy \citep{MHK03}. It is very probable than IRAS 08339+6517 hosts %%@
%galactic wind phenomena, but our data suggest to rule out that knot B was associated with them.

\item {\bf TDG nature}. Dwarf objects showing high metallicities and decoupled kinematics are characteristics of %%@
candidate tidal dwarf galaxies, TDGs \citep{Du00}. The $B$-magnitude derived for this region is $m_B\sim$ 20.1, %%@
implying an absolute $B$-magnitude of $M_B\sim\ -14.4$. Thus, considering its oxygen abundance, this object is away %%@
from the relation given by \citet{RM95} for dwarf irregular galaxies. Following the analysis performed by %%@
\citet{WDF03}, it means that knot B (perhaps, even all the NW arm, knot \#7 in Figure~\ref{figHa}) could be a TDG %%@
candidate. The mean metallicity of the objects in the \citet{WDF03} sample is 12+log(O/H)=8.34$\pm$0.14, but at least %%@
five have oxygen abundances higher than 8.55 (see their Figure 3), similar to the one derived for knot B. However, a %%@
genuine TDG must be a self-gravitating entity. We can not derive it from our spectrum, so it is not possible to %%@
confirm that B is really a TDG candidate made from material stripped from the internal regions of the galaxy. 

\item {\bf Merger nature}. Knot B could be a remnant of a previous merging process suffered by IRAS 08339+6517 in the %%@
past. It would explain the disturbed morphology in its outer regions, specially the long arc of mate\-rial connecting %%@
the north of the galaxy with the southern bright ray (see Figure~\ref{figR}). This scenario was previously suggested %%@
by \citet{CSK04} and would also account for the intensive star-formation activity throughout the galaxy. If this %%@
early merger really happened, it would be in a very advanced state. Following the detailed analysis of a galactic %%@
merger sequence presented by \citet{HG96}, late-stage mergers have tidal appendages emanating from a single nucleus %%@
surrounded by a mostly relaxed stellar profile. Their deep $R$ image of NGC 3921, prototype of this stage of merging, %%@
has morphological signatures similar to those found in IRAS 08339+6517. Other interesting minor merger is the %%@
\emph{Atoms-for-Peace} galaxy, NGC 7252: \citet{HGGS94} found that most of its \ion{H}{i} mass resides outside the %%@
merger remnant, also similar to the case of IRAS 08339+6517. However, if the merger hypothesis is right, the %%@
interaction with the companion galaxy would not be necessary and both the optical plume and the detected \ion{H}{i} %%@
tidal tail  (see Figure~\ref{compara}) would be aligned to the direction of the companion galaxy only by chance. The %%@
effects of the interaction with the external companion galaxy can not be discarded. Although a minor merger occurred %%@
at early times the interaction with the external object seems to be more important at the present time. 
%However, this assumption does not explain the prominent \ion{H}{i} tidal tail found in the direction of the %%@
%companion galaxy nor the fact that this tail possesses around the 70\% of the neutral gas of the system. %%@
%Furthermore, will the optical plume detected in that direction coincide with the \ion{H}{i} tail (see %%@
%Figure~\ref{compara}) only by chance? 
%The X-ray map published in \citet{SS98} also shows disturbed morphology including a weak tail in its NW region, %%@
%although it should not be interpreted as an interaction-induced feature similar to the \ion{H}{i} tail. 
\end{enumerate}

\subsubsection{The non-AGN nature of IRAS 08339+6517}

IRAS 08339+6517 does not host an active galactic nucleus (AGN). Several facts support this statement:

\begin{enumerate}
\item its spectrum, which is similar to the ones of typical starburst galaxies;
\item the FWHM of the emission lines. For example, the FWHM of \Hb\ and [\ion{O}{iii}] $\lambda$5007 are, corrected %%@
for instrumental broadening, 263 km s$^{-1}$ and 259 km s$^{-1}$, respectively; the typical FWHM median value range %%@
for AGNs is between 350 and 550 km s$^{-1}$ \citep{VG97};  
\item the position of the observational data over the \citet{Do00} diagnostic diagrams, that are consistent with the %%@
loci of typical \ion{H}{ii} regions and not with AGN;
\item its logarithmic ratio of FIR to radio flux density, $q$ = 2.34, that is consistent with the values derived for %%@
normal galaxies, $q$ = 2.3 \citep{C92};
\item and the correlation between FIR and radio emission that satisfied the galaxy, both using the FIR and 1.49 GHz %%@
luminosities \citep{Condon91} or the 1.4 GHz radio continuum and 60 $\mu$m FIR fluxes \citep{YRC01}. 
\end{enumerate}

% see Figure~\ref{dopita}].

\subsubsection{A luminous compact blue galaxy}

The properties observed in IRAS 08339+6517 suggest classifying it as a luminous compact blue galaxy (LCBG). LCBGs are %%@
$\sim L^{\star}$ ($L^{\star} = 1.0 \times 10^{10} L_{\odot}$), $M_B\,<\,-$18.5,  blue ($B-V < 0.6$), high mean %%@
surface brightness within the half-light radius ($SB_{B} <$ 21 mag arcsec$^{-2}$), vigorous starbursting galaxies %%@
with an underlying older stellar population \citep{GJK98}. Considering the $B$ magnitude derived for IRAS 08339+6517, %%@
we found that it satisfies all these characteristics: it has $(B-V)$ = 0.05 $\pm$ 0.08, $M_B$ = $-$21.58 $\pm$ 0.04 %%@
and for 2.6$\arcsec$, its half-light radius in the $B$-band, it possesses a mean surface brightness of $SB_{B} \sim$ %%@
17.1 mag arcsec$^{-2}$. 

LCBGs are not common at low redshifts \citep{GOK03} and their evolution and nature is still discussed: whereas %%@
\citet{KG95} and \citet{GK96} suggested that LCBGs are the progenitors of local low-mass spheroidal or irregular %%@
systems experiencing a strong starburst phase, \citet{PG97} and \citet{HGT01} pointed out that they are the %%@
progenitors of present-day bulges of massive spirals. In the local Universe, LCBGs seem to be more evolved objects %%@
and even they could be the equivalent of the high $z$ Lyman-break galaxies \citep{EP03}, being ideal templates for %%@
studies of galaxy evolution and formation. A recent sample of LCBGs in the local Universe was presented by %%@
\citet{WJS04} who found that, on average, LCBGs show strong star-formation activity, emit detectable radio continuum %%@
flux and have lower metal abundances that those expected from the luminosity-metallicity relation for starbursting %%@
emission-line galaxies given by \citet{MS02}. All these properties are also satisfied by IRAS 08339+6517 [the %%@
\citet{MS02} relation gives an oxygen abundance of $\sim$9.2, much higher than that we derived, 12+log(O/H) = %%@
8.45$\pm$0.10].  

\subsubsection{The nature of the dwarf companion galaxy}

The dwarf companion galaxy follows the relation given by \citet{RM95} between absolute $B$-magnitude and oxygen %%@
abundance for dwarf irregular galaxies, indicating that it is not a TDG formed from material striped from IRAS %%@
08339+6517 but an external and independent dwarf galaxy. The apparent component of solid-body rotation found in %%@
\ion{H}{i}, the indications that the \ion{H}{i} gas of the tidal tail has been probably stripped from the main galaxy %%@
and the lower star-formation activity found in the companion galaxy suggest that the effects of the interaction are %%@
not as intense as in IRAS 08339+6517. 

\section{Conclusions}

We have used deep optical and \Ha\ imagery together deep optical intermediate-resolution spectroscopy to analyze the %%@
morphology, colors, ages, stellar populations, phy\-si\-cal conditions, kinematics and chemical abundances of the %%@
galaxy IRAS 08339+6517 and its dwarf companion galaxy, 2MASX J08380769+6508579. Our data reinforce the results of the %%@
\ion{H}{i} observations performed by \citet{CSK04} that both objects are in interaction and that the \ion{H}{i} tidal %%@
tail seems to be formed mainly by material stripped from IRAS 08339+6517.  

We have obtained the oxygen abundances of both galaxies using empirical calibrations. The O/H and N/O ratios of the %%@
two galaxies are rather similar, suggesting a similar degree of chemical evolution of the system. We have analyzed %%@
the chemistry of the ionized gas using different aperture sizes and found an apparent metallicity gradient between %%@
the nucleus and the external zones of IRAS 08339+6517. The reddening coefficient also shows differences between the %%@
central and external zones.

IRAS 08339+6517 shows important \Ha\ emission in its inner regions. IRAS 08339+6517 is not powered by an AGN but it %%@
is a nuclear starbursting galaxy. We have detected at least two different popu\-lations in the galaxy, the age of the %%@
youngest one being around 4 -- 6 Myr. But the more evolved stellar population, with age older than 100 -- 200 Myr, %%@
fits an exponential intensity profile. A model which combines 85\% of a young (6 Myr) population with 15\% of an old %%@
(140 -- 200 Myr) population can explain both the spectral energy distribution and the H Balmer and \ion{He}{i} %%@
absorption lines observed in our spectrum. Furthermore, the NIR colors suggest that even an older stellar population, %%@
with age upper than 1 -- 2 Gyr, is also present in IRAS 08339+6517. Its companion galaxy is practically dominated by %%@
an old population with age $>$ 250 Myr, although it also hosts a recent $\sim$ 6 Myr starburst in its external areas. %%@
The Keplerian, ionized and cold dust masses derived for the main galaxy seem to be the usual for young starbursts.

Our spectra seem to show weak Wolf-Rayet features in IRAS 08339+6517, but they are located only in the central knot %%@
of the galaxy. However, deeper spectra are needed to confirm the WR nature of this galaxy. We discuss that aperture %%@
effects and localization of the bursts with WR stars could play a fundamental role in the detection of this sort of %%@
massive stars in starburst galaxies.

%The detection of WR signatures in this tidally-disturbed galaxy reinforces the \citet{ME00} suggestion that %%@
%interactions with or between dwarf companion objects can be an important trigger mechanism of the star formation %%@
%activity in starburst galaxies. Deeper spectra are needed to confirm the WR nature of IRAS 08339+6517. ******

We have derived the SFR of IRAS 08339+6517 using multi-wavelength correlations. The \Ha, FIR and radio luminosities %%@
give similar values, around 9.5 \Mo\ yr$^{-1}$, suggesting that the contribution of reddening by dust has been %%@
properly estimated. The high SFR derived using X-ray luminosities implies massive stars formation in the last few %%@
Myr, as seems to be supported by the high rate of supernova explosions and the probable detection of a large number %%@
WR stars. 

Knot B, located in the NW arm of the main galaxy, has the higher metallicity and possesses peculiar kinematics. It %%@
could be a TDG candidate because its metallicity is higher than that expected for a dwarf galaxy with its %%@
$B$-luminosity and because the kinematics of the ionized gas are decoupled from the general kinematic pattern. %%@
However, it could also be a remnant of a previous early merger because of the peculiar morphology of the galaxy and %%@
the existence of the \ion{H}{i} tidal tail. Besides it, at the present time the interaction with the external %%@
companion galaxy is much more prominent and probably the origin of the \ion{H}{i} tidal tail.   

Finally, IRAS 08339+6517 could be classified as a luminous compact blue galaxy (LCBG) because of its color, absolute %%@
magnitude and surface brightness. There are very few local LCBGs nowadays detected but nearly half of them have %%@
optical companions, present disturbed morphologies and/or are clearly interacting \citep{GPW04}. An example of local %%@
LCBG is Mkn~1087 \citep{LSER04b}, that is in interaction with two nearby galaxies and shows bridges and tails that %%@
connect dwarf surrounding objects with the main body. If interactions were the responsible of the activity in LCBGs, %%@
it would indicate that they were perhaps more common at high redshifts, as the hierarchical galaxies formation models %%@
predict [i.e., \citet{KW93,Springel05}]. This fact would also support the idea that interaction with dwarf companion %%@
objects could be an important trigger mechanism of the star formation activity in local starbursts.

%especially in those dwarf galaxies where WR stars are detected. }
%These facts support that interactions were perhaps more common at high redshifts,

\begin{acknowledgements}

We thank M\'onica Rodr\'{\i}guez for her kindly comments on the text. We are very grateful to the anonymous referee %%@
for his/her very valuable comments and discussions that have considerably improved this paper.
We would like to acknowledge Anlaug Amanda Kaas and Eric Stempels for their help with NOT observations. We are %%@
indebted to Ricardo Amorin, Ruben S\'anchez and, specially, to Luzma Cair\'os for their help deriving the surface %%@
brightness profiles of the galaxies. A.R.L-S. thanks John Hibbard and Lourdes Verdes-Montenegro for their valuable %%@
comments about radio observations. This work has been partially funded by the Spanish Ministerio de Ciencia y %%@
Tecnolog\'{\i}a (MCyT) under project AYA2004-07466. This research has made use of the NASA/IPAC Extragalactic %%@
Database (NED) which is operated by the Jet Propulsion Laboratory, California Institute of Technology, under contract %%@
with the National Aeronautics and Space Administration.
\end{acknowledgements}

%% The following command ends your manuscript. LaTeX will ignore any text
%% that appears after it.

\end{document}